\documentclass[twocolumn]{aastex631}
\usepackage{amsmath}
\usepackage{amsthm}

\begin{document}

\title{A Unified Mass-Loading Model for ASKAP J1745-5051: Linking Long-Period Radio Transients to Magnetic Cataclysmic Variables}

\author[0009-0007-6439-6891]{Yejing Zhan}
\affiliation{School of Astronomy and Space Science, Nanjing University, Nanjing 210093, People's Republic of China}
5
\author[0000-0003-4157-7714]{Fa-Yin Wang}
\affiliation{School of Astronomy and Space Science, Nanjing University, Nanjing 210093, People's Republic of China}
\affiliation{Key Laboratory of Modern Astronomy and Astrophysics (Nanjing University), Ministry of Education, Nanjing 210093, People's Republic of China}
\correspondingauthor{Fa-Yin Wang}
\email{fayinwang@nju.edu.cn}



\begin{abstract}
Long-period radio transients (LPTs) in white dwarf binaries require sufficiently tenuous plasma to sustain coherent radio emission, posing a challenge in systems with dense accretion flows. ASKAP J1745-5051 (J1745) combines LPT-like radio pulses recurring approximately every 82 min with magnetic accretion and X-ray modulation at a consistent period. We investigate this coexistence using a unified mass-loading framework for Roche-lobe overflow and a tenuous donor wind in the binary magnetic field. The dense stream supplies the accretion curtain and near-white-dwarf X-ray sources, while the tenuous wind feeds flux tubes of $(1–4)\times10^7\,{\rm cm^{-3}}$, satisfying the local density requirement for electron-cyclotron maser emission. A fiducial geometry places the X-ray maximum near orbital phase $0.886$ and produces a radio-visibility window at phases $0.22–0.33$, consistent with the primary observed pulse window. Geometrical modulation and curtain absorption produce stronger soft-band than hard-band X-ray variability. Long-term phase-connected radio monitoring and simultaneous X-ray observations can test this interpretation through recurring radio phase changes and correlated changes in X-ray modulation. These results extend the proposed connection between detached binary LPTs and magnetic cataclysmic variables (mCVs). J1745 may represent an evolved mCV near the orbital-period minimum, suggesting that LPT-like activity can occur at a late accreting stage.
\end{abstract}



\section{Introduction} \label{sec:intro}
\par Long-period radio transients (LPTs) produce recurrent radio pulses on timescales of minutes to hours. Their intensity, polarization, and occurrence can change between successive
cycles or observing epochs \citep{hurley-walkerLongperiodRadioTransient2023,hurley-walkerRadioTransientUnusually2022,dongCHIMEFastRadio2025,blootStronglyPolarisedRadio2025,wangDetectionXrayEmission2025,hymanPowerfulBurstingRadio2005,mcsweeneyNewLongperiodRadio2025,anumarlapudiASKAPJ144834685644Newly2025,hurley-walker29HrPeriodic2024,deruiterSporadicRadioPulses2025,leeEmissionInterpulses645hperiod2025,pritchardDiscovery36minuteLongperiod2026,rosePeriodicRadioXray2026}. The diversity indicates that current LPTs may not originate from a single physical channel. 

\par Proposed interpretations include both single-star and binary systems. In single-star models, the periodicity is associated with the rotation of a magnetized compact object, such as a slowly rotating neutron star, including a magnetar, or a magnetic white dwarf (WD) \citep{hurley-walkerRadioTransientUnusually2022,katzGLEAMXJ16275955235043White2022,konarEnigmaGLEAMXJ162759552350432023,reaLongperiodRadioPulsars2024}. Binary models instead consider a compact star interacting with a companion, such as an M-dwarf star (MD), through magnetic fields or a supply of plasma. The observed periodicity may follow the orbit, stellar spin, or a beat between them. The confirmed WD--MD binary LPTs currently favour a detached, non-interacting configuration: ILT J1101+5521 and GLEAM-X~J0704-37 have orbital periods of approximately 2.1 and 2.9 hr that match their radio recurrence period \citep{hurley-walker29HrPeriodic2024,deruiterSporadicRadioPulses2025,rodriguezSpectroscopicDetection29hour2025}. These associations establish a binary origin for members of the observed population, while the nature of other LPTs remains open.

\par The high brightness temperatures and strong polarization of LPT pulses point to coherent radio emission. Proposed processes include pulsar-like coherent emission, plasma emission, and electron-cyclotron maser emission (ECME)\citep{cooperRotationalDeathlineRadio2024,yangMagneticWhiteDwarf2026,quPhysicalOriginsLong2026}. The plasma conditions for coherent emission depend on the radiation mechanism and the escaping wave mode \citep{melroseCoherentEmissionMechanisms2017}. For low-harmonic ECME, the source must satisfy a low plasma-to-cyclotron frequency ratio, while propagation to the observer imposes additional constraints. Maintaining such a radio-capable channel alongside a dense accretion flow is therefore a central challenge for an accreting binary.

\par Magnetic cataclysmic variables (mCVs) provide important, but limited, examples of the coexistence of accretion and coherent radio emission. In a VLA survey of 122 mCVs, 33 systems were detected at radio wavelengths, and 24 of those detections showed highly circularly polarized emission lasting a few minutes \citep{barrettRadioObservationsMagnetic2020}. A classic example is the accreting polar AM Her, in which a $\sim10$-min outburst at 4.9~GHz was about 20 times brighter than the quiescent emission and nearly 100\% circularly polarized \citep{dulkRadioEmissionAM1983}. However, these ordinary mCV radio events are intermittent flares, not LPT-like pulses with a stable periodicity. The mCV radio flares are more likely to originate from the ECME in the lower corona of the MD donor due to the stellar activity \citep{barrettRadioObservationsMagnetic2020}.

\par Evolutionary calculations for detached binary LPTs predict that orbital shrinkage will eventually bring the donor into Roche-lobe contact and form a mCV. The magnetic field strength then determines whether the descendant is an
intermediate polar (IP) or a polar \citep{rodriguezWhiteDwarfDwarf2026}.
Thus, detached binary LPTs provide a plausible pre-mCV population  \citep{yangMagneticWhiteDwarf2026}.     

\par Recent work shows that ASKAP~J174508.9-505149 (hereafter J1745), an LPT that shows evidence for accretion, brings together LPTs and mCVs \citep{rosePeriodicRadioXray2026}. Optical spectroscopy identifies J1745 as an accreting mCV. Radial velocities give an orbital period of $1.368\pm0.053~{\rm hr}$. Its radio pulses recur at a period of $1.34497~{\rm hr}$, consistent with the orbital period. The X-ray observations independently support the magnetically channelled accretion scenario \citep{imbrognoXrayEmissionLongperiod2026}. XMM-Newton and Einstein Probe detect a modulation at $4868\pm22~{\rm s}$, consistent with the radio and spectroscopic periods. These components show that J1745 is not merely a binary LPT, but a system in which radio emission coexists with ongoing accretion \citep{imbrognoXrayEmissionLongperiod2026,rosePeriodicRadioXray2026,kniggeLongPeriodRadioTransient2026}.


\par We investigate this possibility with a common mass-loading description for a ballistic stream and a tenuous wind from the donor. In the combined dipolar field of the two stars, the Roche-lobe overflow orbits the central WD, forming a ballistic stream, and is loaded by the magnetic field, following its connected field line toward the WD. The stellar wind is loaded in a different region due to its tenuous density. The dense stream supplies the near-WD shock, producing hard X-ray emission and a heated soft X-ray source, while wind-loaded tubes are expected to produce ECME radio emission. This construction links the two candidate emission regions to their plasma supply within one binary geometry.

\par This paper is organized as follows. Sect. \ref{sec:obs_requirement} summarizes the observational constraints. Sect. \ref{sec:loading_framework} and \ref{sec:result} present the unified mass-loading framework and its X-ray and radio results. Sect. \ref{sec:discussion} discusses the energetics, beat interpretation, and LPT--mCV connection, followed by our conclusions in Sect.~\ref{sec:conclusion}.

\section{Observational Requirements}\label{sec:obs_requirement}

\par The X-ray modulation is comparatively smooth across observations, whereas the radio emission has multicomponent pulses, drifting frequency structure, and intervals of several hours during which it switches off. Moreover, radio bursts are observed at different epochs around both binary conjunctions rather than a single orbital phase \citep{rosePeriodicRadioXray2026,imbrognoXrayEmissionLongperiod2026}. The system is therefore well-constrained by the observations. We describe the observational constraints in the following sections. 

\subsection{X-ray constraints}

\par Folding the XMM-Newton data gives sinusoidal pulsed fractions (PF; defined as the semi-amplitude of the sinusoid divided by the source average count rate) of $28\pm1\%$ in the soft 0.3--1.5~keV band, $17\pm1\%$ in the hard 1.5--10~keV band, and $23\pm1\%$ over 0.3--10~keV, while the Einstein Probe full-band data give $22\pm3\%$ \citep{imbrognoXrayEmissionLongperiod2026}. The hardness ratio reaches its maximum near the flux minimum, which means that the emission is harder at the flux minimum. In the orbital-phase convention of \citet{rosePeriodicRadioXray2026}, the X-ray profile peaks at $\phi_{\rm X}=0.89\pm0.19$, close to one of the two conjunctions.

\par The phase-averaged XMM-Newton spectrum is fitted by ${\tt constant}\!\times{\tt tbabs}\!\times{\tt tbpcf}\!\times{\tt gabs}\!\times({\tt apec}+{\tt bbodyrad}),$ \citep{imbrognoXrayEmissionLongperiod2026}. The Galactic column is fixed at $N_{\rm H,Gal}=1.5\times10^{21}~{\rm cm^{-2}}$, while the local partial-covering component has $N_{\rm H,loc}=0.8^{+0.5}_{-0.2}\times10^{22}~{\rm cm^{-2}}$ with a covering fraction $0.8^{+0.1}_{-0.3}$. The continuum contains a collisionally ionized, optically thin component with $k_{\rm B}T_{\rm hot}=15^{+4}_{-3}~{\rm keV}$ and a blackbody with $k_{\rm B}T_{\rm B}=0.14^{+0.03}_{-0.01}~{\rm keV}$. The fitted spectrum also contains an absorption feature at $0.77\pm0.01~{\rm keV}$, tentatively associated with O~VII \citep{imbrognoXrayEmissionLongperiod2026}.

\subsection{Radio constraints}

\par The radio emission is epoch-dependent. The ASKAP/ATCA bursts cluster near the conjunction at $\phi\simeq0.25$, with a measured median phase $\phi_{\rm A/A}=0.31\pm0.03$, whereas the MeerKAT bursts cluster near the opposite conjunction, with $\phi_{\rm MKT}=0.8\pm0.1$ \citep{rosePeriodicRadioXray2026}. These groups were observed from separate epochs, but not from simultaneous cycles. 

\par The observations also sample different frequency ranges. The ATCA measures the bursts in the 1.1--3.1 GHz band, with the detected pulses occupying approximately 1.5--2.7~GHz, while the MeerKAT observed bursts at 0.856--1.712 GHz, with the reported pulse structure concentrated in the 1.3--1.5 GHz\citep{rosePeriodicRadioXray2026}. The polarization properties likewise vary between the two observing campaigns. Together with their half-cycle phase separation, the differences in detected frequency range and polarization behaviour suggest that the ATCA and MeerKAT detections may be from different, epoch-dependent radio-emission branches.

We use the ASKAP/ATCA group, with median phase $\phi_{\rm A/A}=0.31\pm0.03$, as the primary constraint on the fiducial radio geometry. The MeerKAT group, with $\phi_{\rm MKT}=0.8\pm0.1$, is considered separately as an epoch-dependent state in Sect.~\ref{sec:phase_switching}. With this convention, the primary radio window and the X-ray maximum are separated by $\Delta\phi\sim 0.58$, i.e., approximately half an orbit \citep{rosePeriodicRadioXray2026}.

\section{Unified Mass-loading Framework}\label{sec:loading_framework}
\par In the mass-loading framework, matter leaving the inner Lagrange point (L1) initially follows a ballistic trajectory. Once magnetic stresses dominate over the transverse material hydrodynamic stresses, the flow element is captured by a WD-connected field line and is redirected toward the magnetic poles. 


\par The coexistence of accretion-powered X-rays and coherent radio emission requires diverse plasma densities in different tubes that form different emission regions. We follow each source element from its pre-loading trajectory to its first magnetic-loading point and then along the connected field line to the WD. This construction results in distinct emission regions due to the different densities of the plasma sources.

\par We consider a magnetized WD with a mass of $M_1$, a radius of $R_1$, and a magnetic dipole moment of $\boldsymbol\mu_1$, together with a magnetized late-type donor of mass $M_2$, radius $R_2$, and a magnetic dipole moment of $\boldsymbol\mu_2$. The WD is at $\boldsymbol{x}_1=\boldsymbol{0}$ and the donor is at $\boldsymbol{x}_2=a\hat{\boldsymbol{x}}$ in a frame corotating with the binary,
where $a$ is the separation. For a circular orbit, 
\begin{equation}
    \boldsymbol{\Omega}_{\rm orb}
    =\Omega_{\rm orb}\hat{\boldsymbol{z}},
    \qquad
    \Omega_{\rm orb}
    =\left[\frac{GM}{a^3}\right]^{1/2}.
    \label{eq:orbital_frequency}
\end{equation}
where the separation $a=(GMP_\text{orb}^2/4\pi^2)^{1/3}$ with the orbital period $P_\text{orb}$ and total mass $M=M_1+M_2$. At a field point $\boldsymbol{x}$, we define $\boldsymbol{r}_i=\boldsymbol{x}-\boldsymbol{x}_i$, $r_i=|\boldsymbol{r}_i|$, and $\hat{\boldsymbol{r}}_i=\boldsymbol{r}_i/r_i$ where $i=1,2$. In Gaussian cgs units, the field of a centered dipole is
\begin{equation}
    \boldsymbol{B}_i(\boldsymbol{x},t)=\frac{3\hat{\boldsymbol{r}}_i[\boldsymbol{\mu}_i(t)\cdot\hat{\boldsymbol{r}}_i]-\boldsymbol{\mu}_i(t)}{r_i^3}.
    \label{eq:joint_dipole_field}
\end{equation}
The joint field is therefore $\boldsymbol{B}_{\rm tot}=\boldsymbol{B}_1+\boldsymbol{B}_2$.

\par The calculation uses this joint vector field throughout. The flow is captured by the magnetic field at the loading point, while the captured flow follows the magnetic field line, calculated by 

\begin{equation}
    \frac{\mathrm d\boldsymbol{x}}{\mathrm d\ell}
    =\pm\frac{\boldsymbol{B}_{\rm tot}}{|\boldsymbol{B}_{\rm tot}|}
    \label{eq:field_line_tracing}
\end{equation}
Deformation by the loaded plasma and magnetic reconnection are not included in this work.

\par
\begin{figure*}[htbp]
    \centering
    \includegraphics[width=\linewidth]{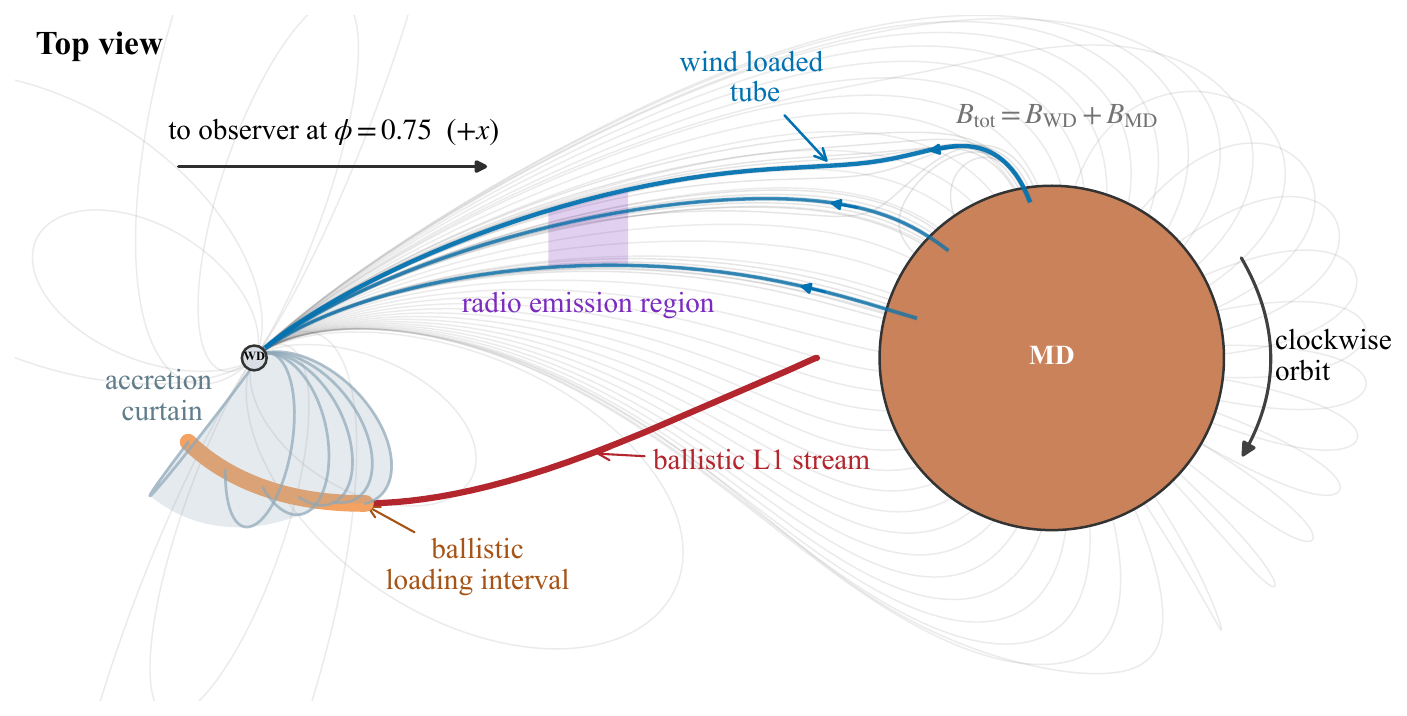}
    \caption{Top-view schematic of the unified mass-loading geometry. In the corotating orbital plane, a finite-width ballistic stream (red curve) leaves L1 and is captured by the magnetosphere over the loading interval (orange thick curve) when the magnetic stress of the joint field $\boldsymbol{B}_{\rm tot}=\boldsymbol{B}_{\rm WD}+\boldsymbol{B}_{\rm MD}$ first exceeds the transverse hydrodynamic stress. The captured stream subsequently follows WD-connected field lines and forms a dense, high-$\Lambda$ accretion curtain (shaded blue), where $\Lambda=\mathrm{d}\dot{M}/\mathrm{d}\Phi$. In parallel, tenuous wind material from the MD is loaded directly onto different WD-connected flux tubes (dark blue curves), producing a low-$\Lambda$ branch that can satisfy the density requirement for ECME. The light-purple region illustrates the schematic radio-emission region. Pale curves denote representative joint magnetic field lines. The observer direction at $\phi=0.75$ and the clockwise orbital motion of the donor are indicated.}
    \label{fig:ballistic_mag_tube}
\end{figure*}

 Before being captured, dense accretion flow is treated as a finite-width ballistic flow launched near $L_1$, demonstrated in Fig. \ref{fig:ballistic_mag_tube}. We use $(x,y,0)$ for the flow centerline and denote $\mu_b=M_2/M$. The trajectory of the flow is controlled by gravitational force, centrifugal force, and Coriolis force in the co-rotational frame.
\par Given the definition of the Coriolis acceleration, $\boldsymbol a_\text{cor}=2\boldsymbol v\times \boldsymbol\Omega_\text{orb}$, the Coriolis component is $\boldsymbol a_\text{cor}=(2\dot y\Omega_\text{orb},-2\dot x\Omega_\text{orb},0)$. 
\par In the co-rotational frame, the centrifugal acceleration points away from the mass center of the binary at $(\mu_b a, 0,0)$. Then,
\begin{equation}
    \boldsymbol a_\text{cen}=\frac{GM}{a^3}(x-\mu_ba, y,0)
\end{equation}

\par The gravitational acceleration is described by 
\begin{equation}
    \boldsymbol a_\text{g}=\frac{GM_i}{d_i^3} \boldsymbol d_i
\end{equation}
where $\boldsymbol r_i=\boldsymbol x-\boldsymbol x_i$ points from from star $i$ to the flow element.

Therefore, the equations of motion are
\begin{equation}
    \begin{aligned}
        \ddot{x}&=2\dot{y}\Omega_\text{orb}+\frac{GM}{a^3}(x-\mu_ba)-\frac{GM_1}{d_1^3}x-\frac{GM_2}{d_2^3}(x-a),\\
        \ddot{y}&=-2\dot{x}\Omega_\text{orb}+\frac{GM}{a^3}y-\frac{GM_1}{d_1^3}y-\frac{GM_2}{d_2^3}y.
    \end{aligned}
\end{equation}
\par The accretion flow would be partially stripped and loaded onto the local magnetic field line once the local field and the density meet the capture condition. Different loading footpoints naturally separate the emission regions. 

The sound speed at the L1 nozzle is
\begin{equation}
    c_s=\left(\frac{k_{\rm B}T}{\mu_gm_{\rm p}}\right)^{1/2},
    \label{eq:stream_sound_speed}
\end{equation}
where $\mu_p$ is the proton mass. We adopt an L1 gas temperature $T=5000~{\rm K}$ and mean molecular weight $\mu_g=1.30$, giving $c_s=5.63~{\rm km\,s^{-1}}$. We adopt the Lubow--Shu transverse scale and an effective stream area \citep{lubowGasDynamicsSemidetached1975},
\begin{equation}
    H_{L1}=f_H\frac{c_s}{\Omega_{\rm orb}},\qquad A_{\rm stream}=f_A\pi H_{L1}^2,
    \label{eq:stream_cross_section}
\end{equation} 
where $f_H=0.5$ and $f_A=1$ allow for the uncertain nozzle width and cross-sectional normalization. The centerline density is then 
 \begin{equation}
    \rho_0(x,y)=\frac{\dot M}{A_\text{stream}v(x,y)}
\end{equation}
where $v=\sqrt{\dot x^2+\dot y^2}$ is the stream velocity.
\par The finite extent of the accretion flow is specified by mass fraction $q$. For a chosen value of $q$, the density entering the loading condition is  
\begin{equation}
    \rho_q(x,y)=q\rho_0(x,y).
    \label{eq:curtain_boundary_density}
\end{equation}

We adopt two values, $q=q_{\rm first}$ and $q=q_{\rm core}=1$ to determine the two first-loading solutions that bound the loading region. The $q<q_\text{\rm first}$ part is treated as the tenuous outer tail of the flow and is ignored here.

\par The WD magnetic field captures the accretion flow if the magnetic pressure dominates the flow's hydrodynamic pressure. We adopt the magnetic-stress scale $B^2/4\pi$ and locate the first loading point at which
\begin{equation}
    \frac{|\boldsymbol B_\text{tot}|^2}{4\pi}\geq \xi \rho(v_\perp^2+c_s^2)
    \label{eq:loading_condition}
\end{equation}
where $v_\perp$ is the velocity perpendicular to the local field line, and $\xi$ is the geometry correction factor. 

\par After a stream element reaches its first magnetic-loading point, it is constrained to move along the corresponding field line toward the WD. We unify the mass-loading process using the mass-loading rate per unit magnetic flux $\Lambda$
\begin{equation}
    \Lambda=\frac{\mathrm d \dot M}{\mathrm d \Phi}
\end{equation}
This rate represents how dense the plasma is in the magnetic tubes. For a ballistic element at the loading point,
\begin{equation}
    v_{\rm load}=\left(v_\perp^2+c_s^2\right)^{1/2},
    \qquad
    \Lambda=\frac{\rho_{\rm load}v_{\rm load}}{B_{\rm load}}.
    \label{eq:ballistic_lambda}
\end{equation}
The field-aligned speed after loading is then
\begin{equation}
    v_\parallel^2(\ell)
    =v_{\parallel,\rm load}^2
    +2GM_1\left[\frac{1}{r(\ell)}-\frac{1}{r_{\rm load}}\right].
    \label{eq:field_aligned_velocity}
\end{equation}
where $\ell$ denotes the arc length measured along a loaded magnetic field line, with $\ell=0$ at the first magnetic-loading point and increasing toward the WD surface.  
Magnetic-flux conservation and mass continuity indicate that $\Lambda$ is conserved along the tube, which gives
\begin{equation}
    B(\ell)A(\ell)=\mathrm d\Phi,
    \qquad
    \rho(\ell)v_\parallel(\ell)A(\ell)=\mathrm d\dot{M},
    \label{eq:tube_conservation}
\end{equation}
and hence
\begin{equation}
    \rho(\ell)=\frac{\Lambda B(\ell)}{v_\parallel(\ell)},
    \qquad
    n_{\rm e}(\ell)=
    \frac{\Lambda B(\ell)}
    {\mu_{\rm e}m_{\rm p}v_\parallel(\ell)}.
    \label{eq:tube_density}
\end{equation}
where $\mu_{\rm e}\simeq 1.18$ is the mean molecular weight per free electron. Thus, the ballistic stream and the two $q$-selected first-loading solutions fix the boundaries of the loaded region, whereas the connected magnetic flux and subsequent field-aligned acceleration fix its footprints and downstream density.

\subsection{X-ray Emission from the High-loading Branch}
\par We assume that the X-ray emission region is formed by the post-loading flow near the WD of the ballistic flow. The dense ballistic flow naturally populates the high-$\Lambda$ curtain, linking the loading point and the WD footprint, illustrated in Fig. \ref{fig:ballistic_mag_tube}. In this branch, radio emission is highly suppressed because the plasma density is too high for a radio-emitting environment and for the emission to escape. 

\par The matter falling onto the WD surface releases gravitational potential energy. The available accretion luminosity is approximately 
\begin{equation}
    L_{\rm acc}\simeq \eta_\text{X}\frac{GM_{1}\dot M}{R_{1}}
    \label{eq:Xray-luminosity}
\end{equation}
where $\eta_\text{X}$ is the energy transfer efficiency. Due to magnetic flux conservation and gravitational acceleration, the matter becomes denser and moves faster, even becoming supersonic, as it moves along the magnetic tube. The supersonic flow cannot settle continuously into the WD atmosphere and therefore forms a standing shock above the magnetic footprint at $r_\text{sh}$ \citep{aizuXRayEmissionRegion1973,cropperEffectsGravityStructure1999}.
For a strong, non-relativistic shock in a monatomic gas, the jump conditions are
\begin{equation}
    v_{\rm sh}=\frac{v_0(r_{\text{sh}})}{4},
    \qquad
    \rho_{\rm sh}=4\rho_0(r_{\text{sh}}).
    \label{eq:shock_jump}
\end{equation}
For a small $r_{\rm sh}\sim R_1$, the corresponding immediate strong-shock temperature is
\begin{equation}
\begin{split}
     k_{\rm B}T_{\rm im, sh}&=\frac{3}{16}\mu_{\rm sh} m_{\rm p}v_0^2\\
     &\simeq\frac{3}{8}\mu_\text{sh} m_{\rm p}GM_1\left(\frac{1}{R_1}-\frac{1}{r_{\rm load}}\right),
\end{split}
    \label{eq:shock_temperature}
\end{equation}
where $\mu_{\rm sh}\equiv\rho/m_{\rm p}(n_{\rm e}+n_{\rm i})\simeq 0.62$ is the mean mass per thermal particle in units of the proton mass in the shock \citep{mukaiXRayEmissionsAccreting2017}. Here $\rho$ is the mass density, and $n_{\rm e}$ and $n_{\rm i}$ are the electron and ion number densities, respectively. The second expression neglects the kinetic term at the loading point. 

\par The shock height $h_\text{sh}=r_\text{sh}-R_1$ is estimated from the radiative cooling length. The specific mass flux at the shock is 
\begin{equation}
    \dot{m}_{\rm sh}\equiv\frac{\mathrm d\dot{M}}{\mathrm dA_{\rm sh}}=\Lambda B_{\rm sh}.
    \label{eq:specific_mass_flux_lambda}
\end{equation}
For a bremsstrahlung-dominated strong shock, the local cooling time of the shock can be estimated by \citep{mukaiXRayEmissionsAccreting2017,vanboxsomNumericalSimulationsHighenergy2018}
\begin{equation}
    t_{\rm cool,sh}=\frac{
        \frac{3}{2}(n_{\rm e,sh}+n_{\rm i,sh})k_{\rm B}T_{\rm im, sh}
    }{
        n_{\rm e,sh}n_{\rm i,sh}\mathcal{L}_{\rm cool}(T_{\rm im,sh},Z)
    }\simeq \frac{3k_{\rm B}T_{\rm im, sh}m_\text{p}v_\text{sh}
    }{\dot m_\text{sh}\mathcal{L}_{\rm cool}(T_{\rm im, sh},Z)
    }
    \label{eq:shock_cooling_time}
\end{equation}
where $\mathcal{L}_{\rm cool}(T_{\rm im, sh},Z)$ is the optical thin cooling function and $n_{\text{i,sh}}\sim n_{\text{e,sh}}\propto \dot m$ are the electron and ion density in the shock.

The characteristic height is then
\begin{equation}
    h_{\rm sh}\simeq v_{\rm sh}t_{\rm cool,sh},
    \label{eq:shock_height}
\end{equation}
These relations provide the physical estimate of the shock scale. In the geometric light-curve calculation, the adopted $h_{\rm sh}$ is used to remove the hot near-WD segment from the cold curtain, resulting in an optically thin and unabsorbed region for soft emission propagation. This would be discussed in Sect. \ref{sec:modulation}. A full radiative-hydrodynamic post-shock solution is not evaluated there.

%

\par The hard X-ray component originates in the optically thin post-shock column, which should contain bremsstrahlung continuum and atomic emission of the ionized plasma. The line emission is ignored in this framework for simplicity. For the broadband calculation, the normalized continuum is represented by
\begin{equation}
    S_{\rm hard}(E)=A_\text{h}\exp\left(-\frac{E}{k_BT_{\rm sh}}\right),
    \label{eq:hard_spectral_shape}
\end{equation}
where $A_\text{h}$ is the normalization factor and $T_{\rm sh}$ is an effective temperature for the multi-temperature
post-shock column.

\begin{figure}[h]
    \centering
    \includegraphics[width=\linewidth]{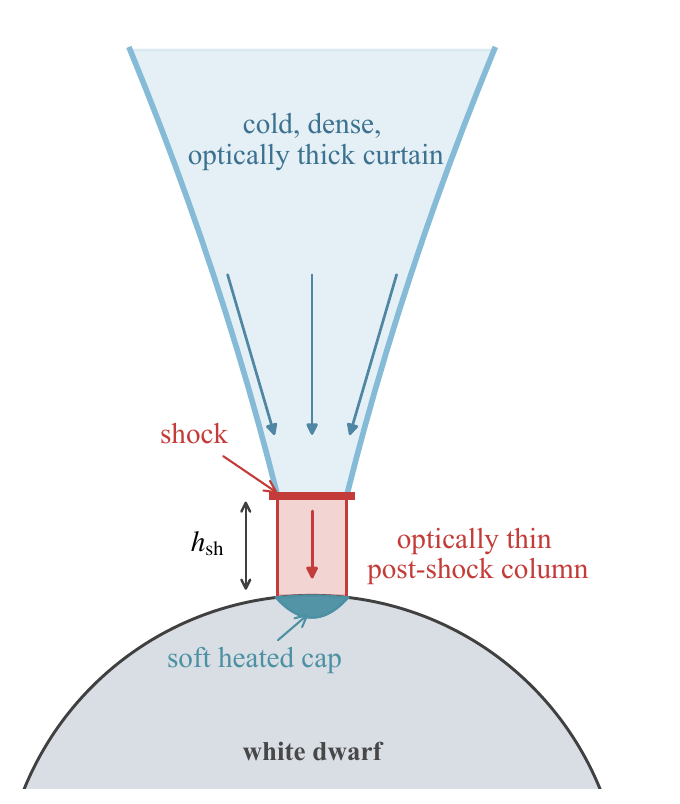}
    \caption{Schematic illustration of the X-ray emission geometry. Once the ballistic accretion stream couples to the WD magnetosphere, the material is channeled along magnetic field lines toward the WD surface, forming an accretion curtain. The flow becomes supersonic and develops a standing shock above the magnetic footprint at $r_{\rm sh}=R_1+h_{\rm sh}$.  Radiative cooling of the optically thin post-shock column produces predominantly thermal bremsstrahlung hard X-rays. Downward irradiation, thermal conduction, and direct thermalization heat this magnetic footprint and produce an optically thick soft cap, emitting the soft component.}
    \label{fig:XrayMechanism}
\end{figure}

\par A fraction of the accretion power is deposited in the WD atmosphere beneath the post-shock column. Downward irradiation, thermal conduction, and the direct thermalization of dense material can heat this magnetic footprint and produce an optically thick soft component\citep{kingAccretionHighlyMagnetized1979,lambUVRadiationAccreting1979}. Details are discussed in Sect. \ref{sec:xray_energetics}.

The soft component is represented by a normalized blackbody spectral shape,
\begin{equation}
    S_{\rm soft}(E)=A_\text{s}\frac{E^3}{\exp(E/kT_{\rm cap})-1}.
    \label{eq:soft_spectral_shape}
\end{equation}
where $A_\text{s}$ is the normalization factor and $T_\text{cap}$ is the effective temperature of the soft cap.  The hard and soft components are normalized as complementary channels of the same accretion-power budget within the adopted 0.5--10~keV reference band.
The high-loading branch therefore provides a common origin for the post-shock hard-X-ray column and the heated soft-X-ray footprint, with their densities, energy fluxes, and characteristic length scales all tied to $\Lambda$.

\subsubsection{Angular Modulation of the X-ray Emission} \label{sec:modulation}
\par The X-ray light curve is modulated by the orbital motion due to the curtain covering and WD self-occultation \citep{nortonSpinModulatedXray1989,mukaiXRayEmissionsAccreting2017}.  For the soft emission, we describe its angular modulation in the tiny-patch-source limit at the WD endpoint of the dense-core ($q=1$) field line, namely the soft cap. The angle modulation is determined by the angle between the normal vector of the cap patch and the observer \citep{evansWhyIntermediatePolars2007}. For the observer at $\hat{\boldsymbol{o}}(i,\phi)$ with inclination $i$ and phase $\phi$, and the normal cap vector $\hat{\boldsymbol n}_{\rm cap}$ with colatitude $\beta_{\rm cap}$ and facing phase $\phi_\text{cap}$, the angle between these two vectors is given by
\begin{equation}
\begin{aligned}
    \mu(\phi)
    &\equiv\hat{\boldsymbol n}_{\rm cap}\cdot
    \hat{\boldsymbol o}(\phi) \\
    &=\cos i\cos\beta_{\rm cap}
    +\sin i\sin\beta_{\rm cap}
    \cos\!\left[2\pi(\phi-\phi_{\rm cap})\right],
\end{aligned}
    \label{eq:xray_direction_cosine}
\end{equation}
With $\mu_+=\max(\mu,0)$, the baseline soft emission modulation factor is 
\begin{equation}
   \mathcal  V_\text{s}(\phi)=\frac{\mu_+(\phi)}{\langle \mu_+\rangle_{\phi}}
    \label{eq:soft_modulation_factor}
\end{equation}
where angle brackets denote a full-cycle phase average. The normalization sets the mean of each factor to unity, so that its amplitude does not redefine the adopted mean luminosity. For non-negative $\mu(\phi)$, the modulation factor can be simplified to $\mathcal V_\text{s}(\phi)=1+\tan i\tan \beta_\text{cap} \cos\left[2\pi(\phi-\phi_{\rm cap})\right]$. 

\par The hard component originates in the optically thin post-shock column immediately above the soft cap. The two emitting regions therefore share a common magnetic footprint and a similar viewing geometry throughout the orbital cycle. Motivated by this spatial association, we adopt a phenomenological sinusoidal modulation for the hard component, with the same phase as the soft-cap modulation,

\begin{equation}
    \mathcal V_\text{h}(\phi)=1+a_\text{h}\cos[2\pi(\phi-\phi_\text{cap})]
    \label{eq:hard_modulation_factor}
\end{equation}
where $a_\text{h}$ is the hard-modulation factor. We adopt $a_h=a_s=\tan i\tan\beta_{\rm cap}$ as a phenomenological reference, aligning the phase and intrinsic fractional modulation of the hard component with those of the soft cap. The assumption is motivated by by the relative locations of the two emitting regions.

\par Except for the geometric modulation, the emission is also modulated by the phase-dependent curtain absorption. Field lines launched from the ballistic footpoint define a three-dimensional curtain sheet, from which the hot segment below the adopted shock height is removed from the curtain absorption. The remaining surface is triangulated, and a ray is traced from the soft-source position toward the observer. We define the intersection function $C_{\rm curt}(\phi)$ to be unity when this ray intersects the cold curtain and zero otherwise.

\par The absorbing column is tied to the density obtained from the mass-loading calculation. A reference point $\boldsymbol{x}_{\rm ref}$ is defined as the first intersection of the source-observer ray with the curtain measured from the loaded ballistic midpoint. The intersection fixes $q_{\rm ref}=0.5$ and $r_{\rm ref}=|\boldsymbol{x}_{\rm ref}|$. The density $n_{\rm e,ref}$ is obtained by logarithmic interpolation in radius along the two ballistic tubes bracketing $q_{\rm ref}$, followed by interpolation in $q$. This evaluates the mass-loading relation $n_{\rm e}(\ell_{\text{ref}})$ and $\rho(\ell_{\rm ref})$, according to Eq.~\eqref{eq:tube_density}. The corresponding hydrogen-nucleus density is $n_{\rm H,ref}=\rho_{\rm ref}/\mu_{\rm H}m_{\rm p}$ and effective local column are therefore
\begin{equation}
    N_{\rm H,eff}=n_{\rm H,ref}L_{\rm eff}
    =\frac{\rho_{\rm ref}}{\mu_{\rm H}m_{\rm p}}L_{\rm eff},
    \label{eq:xray_effective_column}
\end{equation}
where $\mu_{\rm H}=1.4$ is the adopted mean mass per hydrogen nucleus in units of $m_{\rm p}$ and $L_{\rm eff}$ is an effective penetration distance. It is a free parameter representing transverse structure in the cold curtain and is not identified with the physical distance through which the ray penetrates the curtain.
intersection. 
\par The soft-component transmission is
\begin{equation}
\begin{aligned}
    \mathcal T_{\rm s}(E,\phi)
    &=\exp[-N_{\rm H,Gal}\sigma_{\rm cold}(E)]\\
    &\times\left\{1-C_{\rm curt}(\phi)
    +C_{\rm curt}(\phi)\exp[-N_{\rm H,eff}\sigma_{\rm cold}(E)]\right\}
\end{aligned}
    \label{eq:xray_effective_transmission}
\end{equation}
where $N_{\rm H,Gal}$ is the foreground column and $\sigma_{\rm cold}(E)$ is the adopted broad-band photoelectric cross section per hydrogen atom \citep{wilmsAbsorptionXRaysTheInterstellar2000}. If the source ray does not intersect the curtain, $C(\phi)=0$ and the local-transfer factor is unity. Local curtain attenuation of the hard flux is neglected because the adopted curtain column is optically thin for the hard component. The hard component is therefore assigned foreground absorption only,
\begin{equation}
    \mathcal T_{\rm h}(E)
    =\exp[-N_{\rm H,Gal}\sigma_{\rm cold}(E)].
    \label{eq:xray_hard_transmission}
\end{equation}

\par The point-source projection factor and the curtain transmission act simultaneously: $\mathcal V_{\rm s}$ and $\mathcal V_{\rm h}$ set the intrinsic geometric modulation, while $\mathcal T_{\rm s}$ and $\mathcal T_{\rm h}$ apply additional absorption along intersecting rays. Combining with the spectral shapes in Eqs~\eqref{eq:hard_spectral_shape} and \eqref{eq:soft_spectral_shape}, the resulting energy-flux density is
\begin{equation}
    \begin{aligned}
        F_E(\phi,E)=&\frac{L_{\rm acc}}{4\pi d^2}
        \Big[f_{\rm s}S_{\rm soft}(E)\mathcal V_{\rm s}(\phi)\mathcal T_{\rm s}(E,\phi)\\
        &+(1-f_{\rm s})S_{\rm hard}(E)\mathcal V_{\rm h}(\phi)\mathcal T_{\rm h}(E)\Big]
    \end{aligned}
    \label{eq:xray_modulated_spectrum}
\end{equation}
where $d$ is the distance and $f_{\rm s}$ is the soft fraction of the reference-band luminosity, which is defined by the observed best-fit model \citep{imbrognoXrayEmissionLongperiod2026},
\begin{equation}
    f_{\rm soft}=\frac{\int_{0.5}^{10}F_{\text{bbodyrad}}\, \mathrm dE}{\int_{0.5}^{10}\left(F_{\rm bbodyrad}+F_{\rm apec}\right) \mathrm dE}.
    \label{eq:fsoft}
\end{equation}
No additional curtain-reradiation component is included. The phase-dependent photon light curve is defined by 
\begin{equation}
    P(\phi)=\int ^{E_2}_{E_1}\frac{F(\phi,E)}{E}\mathrm d E
    \label{eq:Xray-lightcurve}
\end{equation}
where $[E_1, E_2]$ is the chosen energy band.

\subsection{Coherent Radio Emission from the Low-loading Branch}

\par The donor wind launched from the WD-facing hemisphere supplies the low-loading branch. Its small mass flux per surface element gives $\Lambda$ well below that of the ballistic stream. In the adopted geometry, wind-fed tubes connect the donor to the WD in a distinct region from the dense stream-fed curtain, providing candidate low-density sites for coherent radio emission.

\subsubsection{Electron-Cyclotron Maser Emission}
\par We interpret the coherent radio emission as ECME. The radio emission frequency is determined by the gyromagnetic resonance condition \citep{wuTheoryTerrestrialKilometric1979}:
\begin{equation}
    \omega_\text{ECME}-\frac{s\omega_B}{\gamma}-k_\parallel v_\parallel=0
    \label{eq:resonance_condition}
\end{equation}
where $\omega_B=eB_\text{tot}/m_e c$ is the electron-cyclotron frequency, $s$ is the harmonic number, $\gamma$ is the electron Lorentz factor, and $k_\parallel$ is the wave vector along the magnetic field. The detailed growth, saturation, and mode selection are not calculated here; the model instead tests whether the local plasma and magnetic geometry provide conditions compatible with ECME.

\par A loss-cone distribution can be produced when electrons propagate into the converging field toward the WD. Conservation of the first adiabatic invariant gives \citep{wuElectricallyPoweredBinary2002}
\begin{equation}
    \sin^2\alpha_{\rm lc}(\ell)
    =\frac{B(\ell)}{B_{\rm m}},
    \label{eq:loss_cone_angle}
\end{equation}
where $\alpha$ is the pitch angle and $B_m$ is the field strength at the loss-cone region. Electrons with sufficiently small pitch angles reach the dense WD footpoint and are removed, while those at larger pitch angles mirror before reaching the surface. Continued injection by the donor wind can therefore maintain a deficit at small pitch angles and a positive perpendicular gradient near the loss-cone boundary, providing the free energy required by the maser \citep{wuTheoryTerrestrialKilometric1979, melroseCoherentEmissionMechanisms2017}. Calculating the evolution of the electron distribution is out of the scope of this work. We therefore treat the loss cone as a physically motivated condition rather than a calculated instability.

\par The density along each accepted wind-loaded tube is inherited from the common mass-loading calculation, $n_{\rm e}(\ell)=\Lambda B(\ell)/\mu_{\rm e}m_{\rm p}v_\parallel(\ell)$. With the plasma oscillation angular frequency,
\begin{equation}
    \omega_{\rm p}=\left(\frac{4\pi e^2n_{\rm e}}{m_{\rm e}}\right)^{1/2},
    \label{eq:radio_characteristic_frequencies}
\end{equation}
the ECME requires a sufficiently low plasma density, typically $\omega_\text{p}/\omega_\text{B}\leq 0.1-0.3$ \citep{willesElectroncyclotronMaserEmission2004,treumannElectronCyclotronMaser2006}. Because $n_{\rm e}\propto\Lambda$, these conditions connect the survival of a radio-capable region directly to the low mass loading of the wind-fed tubes.

\subsubsection{Radio visibility}
\par Eq. \ref{eq:radio_characteristic_frequencies} constrains the magnetic field, and then the candidate emission region. Ignoring the parallel Doppler term, the frequency mapping between the observation frequency and the ECME frequency is 
\begin{equation}
    \nu_\text{obs}=\frac{s\nu_B}{\gamma}=s\frac{eB}{2\gamma\pi m_e c}
    \label{eq:obs_freq_mapping}
\end{equation}
For a fractional bandwidth $\delta_\nu$, the constrained magnetic field should be 
\begin{equation}
    B_-\leq B_0=\frac{2\pi \gamma m_{\rm e}c}{e}\frac{\nu_{\rm obs}}{s}\leq B_+
    \label{eq:radio_central_field}
\end{equation}
where $B_\pm=B_0\left(1\pm\delta_\nu/2\right)$ are the field boundaries. The radio calculation uses the same observer position as the X-ray covering calculation. 

We describe the angular pattern by a hollow cone with half-opening angle $\chi$ measured from the local field tangent. For the fiducial geometric calculation, we adopt the approximate parameterization $\cos\chi\simeq\sqrt{1-\gamma^{-2}}$. 

\begin{figure}
    \centering
    \includegraphics[width=\linewidth]{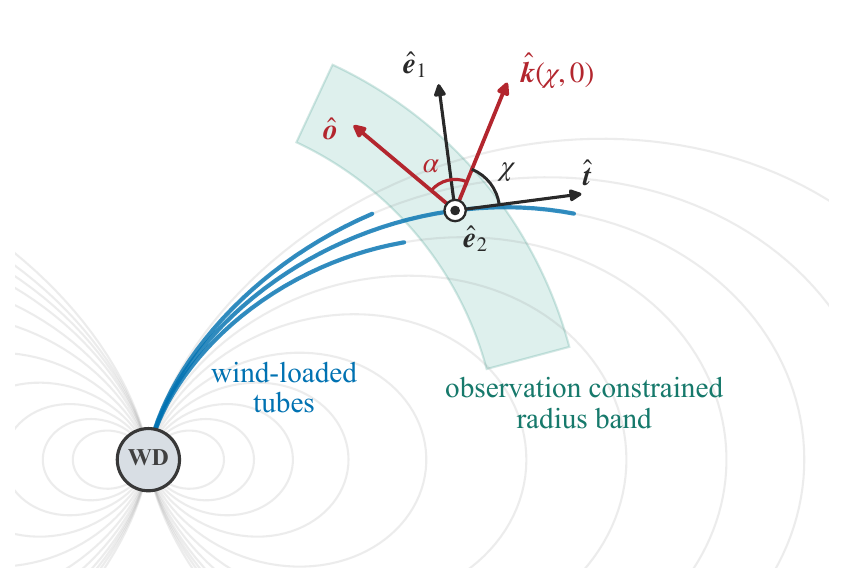}
    \caption{Schematic illustration of the radio emission geometry. The blue curves represent wind-loaded magnetic flux tubes as the candidate ECME emission region. The observed radio frequency constrains the local magnetic-field strength and hence restricts the candidate emission region to the green shaded radial band.
    At each candidate point, one can construct the local transverse basis  $(\hat{\boldsymbol{e}}_1,\hat{\boldsymbol{e}}_2,\hat{\boldsymbol{t}})$. $\hat{\boldsymbol{t}}$ denotes the local field tangent, while $\hat{\boldsymbol{e}}_1$ lies in the $\hat{\boldsymbol{o}}$--$\hat{\boldsymbol{t}}$ plane, and $\hat{\boldsymbol{e}}_2$ is perpendicular to this plane. The emission direction $\hat{\boldsymbol{k}}(\chi,0)$ is specified by the field-emission angle $\chi$, with the emission azimuth chosen to be nearest to the observer. The angular separation $\alpha$ between $\hat{\boldsymbol{k}}(\chi,0)$ and $\hat{\boldsymbol{o}}$ determines the visibility of the radio emission.}
    \label{fig:ECME_geo}
\end{figure}

At each radio-emission candidate point, we denote $\hat{\boldsymbol{t}}=\boldsymbol{B}/B$ as the field tangent. Using the observer direction, one can construct the local transverse basis
\begin{equation}
    \hat{\boldsymbol{e}}_1
    =\frac{\hat{\boldsymbol{o}}
    -(\hat{\boldsymbol{t}}\cdot\hat{\boldsymbol{o}})
    \hat{\boldsymbol{t}}}
    {\left|\hat{\boldsymbol{o}}
    -(\hat{\boldsymbol{t}}\cdot\hat{\boldsymbol{o}})
    \hat{\boldsymbol{t}}\right|},
    \qquad
    \hat{\boldsymbol{e}}_2
    =\hat{\boldsymbol{t}}\times\hat{\boldsymbol{e}}_1.
    \label{eq:ecme_cone_basis}
\end{equation}
where $\hat e_1$ lies in the $\hat{\boldsymbol{o}}-\hat{\boldsymbol{t}}$ plane, while $\hat e_2$ is perpendicular to this plane. Therefore, the direction of the radio-emission hollow cone is 
\begin{equation}
    \hat{\boldsymbol{k}}(\chi,\eta)
    =\cos\chi\,\hat{\boldsymbol{t}}
    +\sin\chi
    \left(\cos\eta\,\hat{\boldsymbol{e}}_1
    +\sin\eta\,\hat{\boldsymbol{e}}_2\right),
    \label{eq:ecme_emission_direction}
\end{equation}
where $\eta$ specifies the azimuth around the cone. The visibility calculation selects the emission directions nearest to the observer, namely $\eta=0$.  The emission-observer angle is therefore

\begin{equation}
    \alpha
    =\cos^{-1}\left(
    \hat{\boldsymbol{o}}\cdot\hat{\boldsymbol{k}}(\chi,0)
    \right)
    \label{eq:radio_beaming_condition}
\end{equation}
For a chosen beaming angle $\theta_b$, the condition $\alpha \leq \theta_b$ implies that the radio emission at this candidate point is visible.  

\section{Result}\label{sec:result}
\par The mass-loading framework developed above provides a common physical distinction between the two candidate emission regions. Each source element follows its initial trajectory to its first magnetic-loading point, then moves along the joint WD--donor magnetic field. Conservation of mass flow per unit magnetic flux, $\Lambda=\mathrm{d}\dot{M}/\mathrm{d}\Phi$, determines the plasma density along the resulting flux tube. The dense, high-$\Lambda$ ballistic stream therefore supplies the accretion curtain, post-shock hard X-ray column, and heated soft X-ray footprint, whereas the low-$\Lambda$ donor-wind tubes provide the low-density environment for ECME.
\par We evaluate a fiducial binary with $M_1=0.83\,M_\odot$, $R_1=6.52\times10^8~{\rm cm}$, $M_2=0.0963\,M_\odot$, $R_2=9.05\times10^9~{\rm cm}$, and $P_{\rm orb}=4868~{\rm s}$, based on the observation\citep{rosePeriodicRadioXray2026,imbrognoXrayEmissionLongperiod2026}. The WD radius follows the adopted mass--radius relation \citep{zalameaWhiteDwarfsStripped2010}, whereas the donor radius is set to its Eggleton Roche-lobe radius \citep{eggletonApproximationsRadiiRoche1983}. The observer is set at $i=15^\circ$. In the co-rotational frame, the observer azimuth is determined by the orbital phase. We use the same binary geometry, combined magnetic field, mass-loading calculation, and phase-dependent observer direction for both the X-ray and radio calculations. 

\par The joint magnetic configuration is set by dipole moments $\mu_{1}=1.5\times 10^{33}~{\rm G~cm}^3$ and $\mu_{2}=7.4\times 10^{31}~{\rm G cm}^3$. The dipole moments result in the surface magnetic field strength of 5.4 MG for the WD and of 100 G for the MD, respectively. The WD dipole has colatitude $\theta_\text{WD}=45^\circ$ and azimuth $\psi_\text{WD}=-50^\circ$, while the donor dipole axis is aligned with the negative orbital axis. This particular configuration is selected as a simultaneous loading-topology solution, rather than adopted as a typical or independently measured field. The parameter scan for dipole colatitude and azimuth is demonstrated in Append. \ref{app:xray_modulation}.

\par As shown in Sect. \ref{sec:obs_requirement}, the soft peak occurs at the orbital conjunction when the soft cap is closer to the observer, indicating a lower optical depth along the line of sight at this phase. For the binary and stream parameters adopted above, the accretion flow remains hydrodynamically dominated beyond L1 and is captured by the magnetic field on the far side of the WD, as illustrated in Fig. \ref{fig:ballistic_mag_tube}. The captured flow forms an accretion curtain that obscures the soft component at other orbital phases through absorption along the line of sight. The donor-wind flow is loaded onto the magnetic field at the MD surface and follows field lines that connect directly to the WD. The robustness of this particular solution is discussed in Sect. \ref{sec:discussion}.


\subsection{X-ray emission} 

\par The X-ray-emitting high-loading branch is supplied by a ballistic stream with $\dot M=5\times 10^{-11}\,M_\odot\,{\rm yr}^{-1}$ from L1. We set the stress coefficient to $\xi=1$ and use $q_{\rm first}=0.01$ together with $q_{\rm core}=1$ to locate the two loading boundaries of the curtain. The resulting loading points determine $\Lambda$, the dense-core footprint, and the density along the loaded tubes.
\begin{figure*}[htbp]
    \centering
    \includegraphics[width=\linewidth]{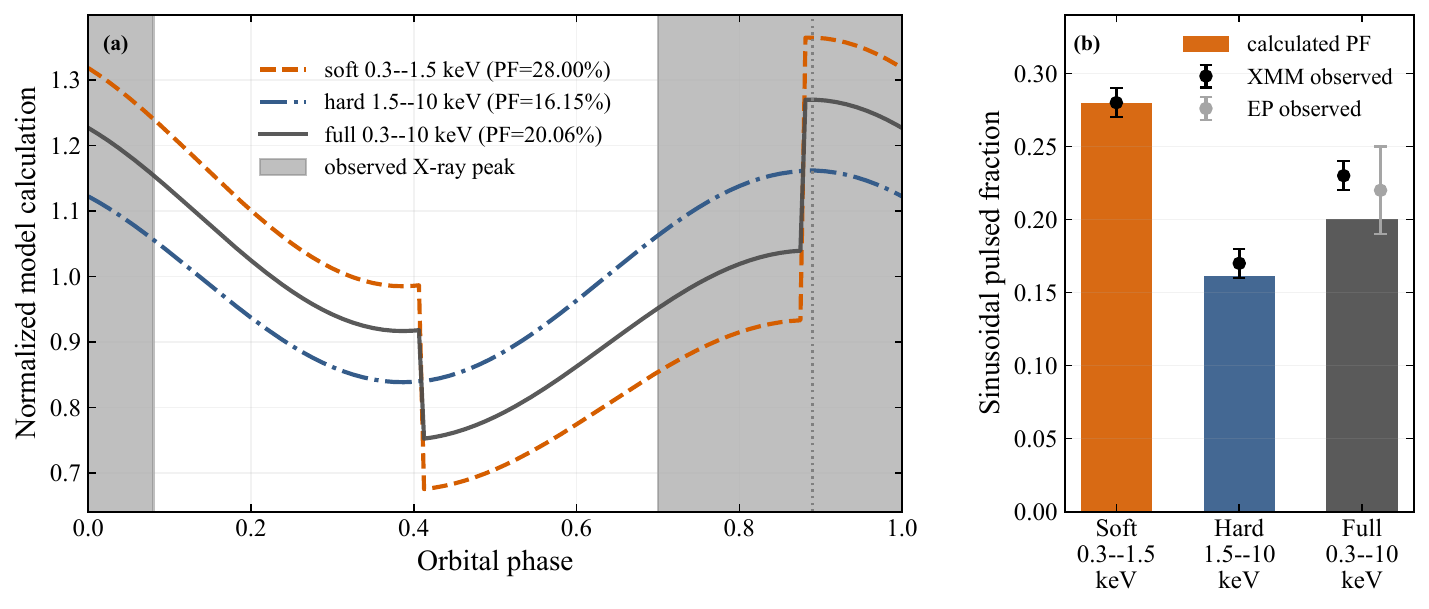}
    \caption{Model calculation of the observed orbital X-ray variability. (a) Orbit-phase-dependent light curves over the soft (0.3–1.5 keV; orange dashed), hard (1.5–10 keV; blue dash-dotted), and full (0.3–10 keV; gray solid) bands. The light curves are normalized. The gray shaded interval marks the observed X-ray maximum at orbital phase $\phi=0.89\pm0.19$ \citep{rosePeriodicRadioXray2026}, with its central phase indicated by the vertical dotted line. The sharp discontinuities in the soft-band curve result from the simplified point-source, curtain-attenuation prescription, in which the line of sight switches abruptly between uncovered and covered states. (b) Calculated sinusoidal pulsed fractions (bars), defined as the fitted fundamental semi-amplitude divided by the mean flux, compared with the XMM-Newton measurements (black points) and the Einstein Probe full-band measurement (gray point) \citep{imbrognoXrayEmissionLongperiod2026}. The model reproduces the observed peak phase and the soft-hard-band modulation.}
    \label{fig:Xray-PF}
\end{figure*}
The two boundaries occur at azimuths $37.70^\circ$ and $128.47^\circ$. The joint field tracing gives an upper-footprint soft cap with colatitude $31.07^\circ$ and an observer phase $\phi_{\rm cap}=0.886$.

\begin{figure}[hbtp]
    \centering
    \includegraphics[width=\linewidth]{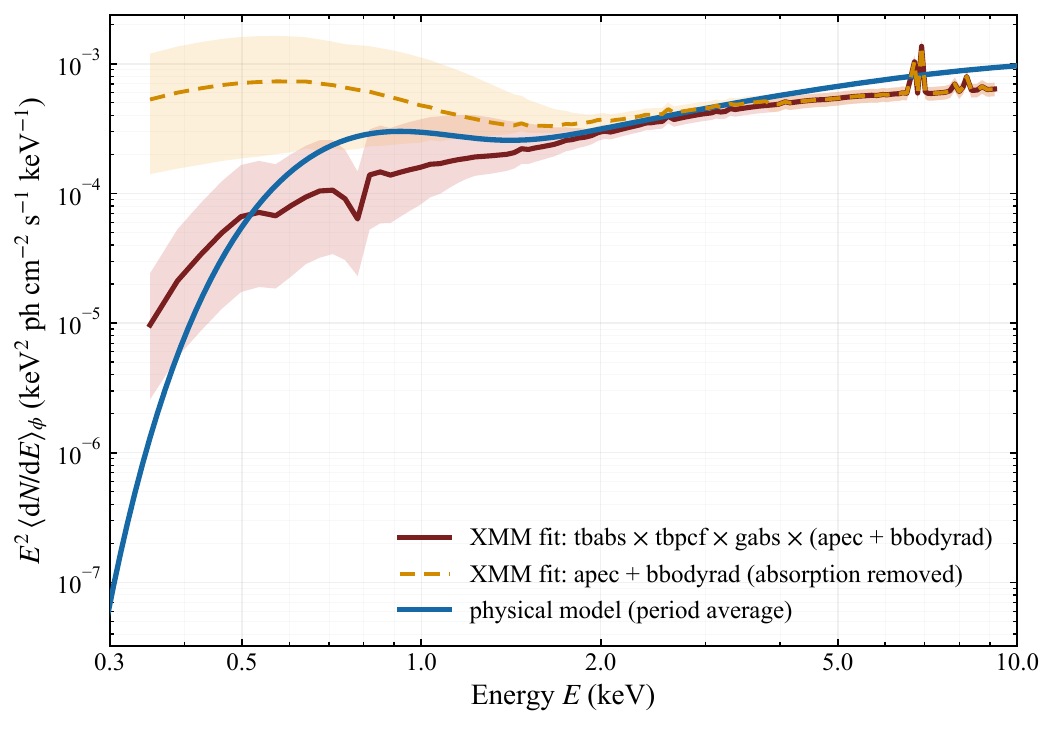}
    \caption{Comparison of the calculated phase-averaged X-ray spectrum with the published XMM-Newton spectral fit. The absorbed best-fitting model, ${\tt tbabs}\times{\tt tbpcf}\times{\tt gabs}\times({\tt apec}+{\tt bbodyrad})$, is shown by the dark-red curve, while the orange dashed curve shows the corresponding intrinsic ${\tt apec}+{\tt bbodyrad}$ continuum. The red and orange shaded regions are the 90\%-confidence-level envelopes. The blue curve is the orbital-phase-averaged spectrum calculated by our physical model, including foreground absorption and the phase-dependent attenuation of the accretion curtain. The physical model reproduces the overall normalization and broad spectral shape of the absorbed XMM spectrum over most of the 0.3–10 keV.}
    \label{fig:XraySpec}
\end{figure}

\par The soft component is emitted from the heated dense-core footprint, namely the soft cap, with $k_{\rm B}T_{\rm cap}=0.14~{\rm keV}$. In the point-source limit, the soft geometrical modulation $\mathcal V_{\rm s}$ is calculated and simplified to $\mathcal V_{\rm s}=1+a_\text{s} \cos[2\pi(\phi-\phi_\text{cap})]$, with $a_\text{s}=0.16145$. 

\par The post-shock continuum is evaluated with $k_{\rm B}T_{\rm sh}=15~{\rm keV}$. Rather than introducing an independent hard-component amplitude, we set its intrinsic modulation to $\mathcal V_h(\phi)=1+a_{\rm s}\cos[2\pi(\phi-\phi_{\rm cap})]$, using both the amplitude and phase obtained from the soft geometrical modulation.

\par The curtain absorption is tied to the intersection function $C_\text{curt}(\phi)$, which is calculated by whether the ray emitted from the cap penetrates the curtain. The function $C_\text{curt}(\phi)$ is unity at phase $\phi\simeq 0.4-0.9$ and zero otherwise. For the cooling-scale estimate in Eq. \eqref{eq:shock_cooling_time}, non-relativistic bremsstrahlung from fully ionized H/He gives $\mathcal{L}_{\rm cool}\simeq4.5\times10^{-23}~{\rm erg\,cm^3\,s^{-1}}$ at a representative immediate post-shock temperature $k_{\rm B}T\simeq35~{\rm keV}$ \citep{rybickiRadiativeProcessesAstrophysics1991}. We remove the hot curtain segment below the fiducial shock height $h_\text{sh}=0.05 R_1$, which is chosen in the range of $0.01-0.11 R_1$ calculated in Eq. \eqref{eq:shock_height} with $q=0.01-1$. We adopt $N_{\rm H,Gal}=1.5\times10^{21}~{\rm cm}^{-2}$ for the foreground column. At the conjunction-phase ray--curtain intersection, we set the reference loading point at $q_\text{ref}=0.5$, corresponding to $\rho_{\rm ref}=4.8\times10^{-10}~{\rm g\,cm^{-3}}$ with $\Lambda=2.3\times 10^{-6} {\rm g\,cm^{-2}\,s^{-1}\,G^{-1}}$. The effective penetration distance $L_\text{eff}=5.2\times 10^6 \text{ cm}$ then gives $N_{\rm H,eff}=10^{21}~{\rm cm}^{-2}$.

\par Both soft and hard emission are normalized. The emission luminosity is defined by Eq. \eqref{eq:Xray-luminosity}. We adopt the energy transfer efficiency $\eta_\text{X}=0.2$. The fraction of the soft and hard components is determined by the fraction $f_s$. We adopt $f_s=0.25$, derived from the observed, unabsorbed, best-fit model, \texttt{apec + bbodyrad} \citep{imbrognoXrayEmissionLongperiod2026}, based on Eq. \eqref{eq:fsoft}. 

\par The calculated orbital light curve for the X-ray emission is demonstrated in Fig. \ref{fig:Xray-PF}, which compares the calculated variability with the XMM-Newton and Einstein Probe measurements. The phase-dependent light curve is integrated over the soft ($0.3$--$1.5~{\rm keV}$; orange dashed), hard ($1.5$--$10~{\rm keV}$; blue dash-dotted), and full ($0.3$--$10~{\rm keV}$; gray solid) bands, according to Eq. \eqref{eq:Xray-lightcurve}. Each resulting light curve is normalized by its mean. Light curves in all three energy bands reach their maximum near $\phi\simeq 0.9$, within the observed X-ray peak interval $\phi=0.89\pm0.19$. The modulation originates from the location of the accretion curtain and the soft cap. Thus, at $\phi_{\rm cap}$, the observer receives the most photons from the cap and the post-shock segment above the cap. The cap--observer ray encounters little curtain material, leading to low absorption. 

\par As the binary rotates away, the angular emission factor decreases. At about $\phi=0.4$, the cap--observer ray intersects the accretion curtain. The curtain then applies strong absorption to the soft emission, leading to a soft emission drop, while the curtain is nearly transparent to the hard component. The hard emission therefore keeps its geometrical modulation. The spectrum consequently becomes harder at the flux minimum: the soft flux decreases by a larger fraction, causing the hard-to-soft count-rate ratio to increase. 

\par The sharp changes in the soft-band curve near $\phi\simeq0.41$ and $0.88$ are a consequence of the point-source, curtain-attenuation prescription. Due to the model simplification, the curtain is treated as equal-thickness, with the effective penetration distance $L_\text{eff}$ and sharply truncated boundaries. The emission ray of the soft cap in the point-source limit penetrates the curtain, switching abruptly between uncovered and covered states. 

\par Fitting the fundamental component of the model light curves gives sinusoidal PF of $28.00\%$, $16.15\%$, and $20.06\%$ in the soft, hard, and full bands, respectively. The soft- and hard-band values closely reproduce the XMM-Newton measurements of $28\pm1\%$ and $17\pm1\%$. The calculated full-band value is lower than the XMM-Newton value of $23\pm1\%$, but is consistent with the Einstein Probe measurement of $22\pm3\%$. 

\par The agreement in the peak phase and energy-dependent modulations demonstrates that the heated-cap geometry, together with curtain absorption, can account for the principal observed X-ray variability of J1745. The fiducial parameters provide one viable configuration for the observed X-ray modulation. Its phase and shape depend on the WD magnetic azimuth and tilt (Append. \ref{app:xray_modulation}).


Fig~\ref{fig:XraySpec} compares the orbital-phase-averaged spectrum calculated by the physical model with the published XMM-Newton best-fitting spectrum from \cite{imbrognoXrayEmissionLongperiod2026}. The best-fitted model, ${\tt tbabs}\times{\tt tbpcf}\times{\tt gabs}\times({\tt apec}+{\tt bbodyrad})$, is shown by the dark-red curve, while the orange dashed curve shows the corresponding intrinsic ${\tt apec}+{\tt bbodyrad}$ continuum. The red and orange shaded regions are the 90\%-confidence-level envelopes. Our model calculation is shown in blue solid curves. The physical model reproduces the broad spectral shape of the absorbed XMM spectrum over most of the $0.3–10~{\rm keV}$ range, particularly the hard continuum above $\sim1~{\rm keV}$. Part of the remaining spectral discrepancy in 0.5-1 keV arises from the simplified treatment of curtain covering. Our simplification of the density gradient and equal-thickness curtain should strongly modify the shade and amplitude of the spectrum in the low-energy band, namely $<1$ keV. The hard post-shock component is treated as much less affected by the local curtain. Thus, the calculated spectrum is more consist to the best-fit spectrum in the hard band. A second difference is that our model does not consider the line absorption by the ionized plasma, which leads to an inconsistency with the best-fit model in the 0.5- 1 keV band, especially the narrow absorption feature near $0.77~{\rm keV}$. 


\subsection{Radio emission}

\par We model the low-loading branch with a donor-wind mass-loss rate $\dot M_\text{w}=10^{-14} M_\odot \text{ yr}^{-1}$ and velocity $v_\text{w}=10 \text{ km s}^{-1}$ at the donor surface, motivated by measurements and constraints on MD winds \citep{vidottoMdwarfStellarWinds2014,woodNewObservationalConstraints2021,waltersMeasurementStellarSubstellar2023}. To generate the tube population, we divide the WD-facing hemisphere of the MD into $N_\alpha\times N_\psi=30\times60$ angular bins, where $\alpha$ measures the angle from the sub-WD direction and $\psi$ is the azimuth about that axis. Each cell is assigned the fraction of the isotropic wind carried by its actual solid angle. 

\par At the adopted parameters, the wind is magnetically dominated already at the MD surface. The resulting mass loading is $\Lambda\simeq (6-23)\times 10^{-12} {~\rm g\,cm^{-2}\,s^{-1}\,G^{-1}}$. The wind-loaded tubes remain strongly magnetically dominated throughout their trajectories from the donor to the WD, with a magnetic-hydrodynamic-pressure ratio over $10^3$. We therefore treat that surface as the first-loading location. The accepted tubes connect the donor directly to the WD. Densities of the tubes are calculated from the same mass-loading relation used for the accretion curtain, namely Eq. \eqref{eq:tube_density}, with no independent radio density introduced.

\par We prescribe the donor-wind launch speed without solving its initial acceleration, while WD gravitational acceleration after loading retains. According to Eq. \eqref{eq:loading_condition}, surface loading requires $\sqrt{v_\perp^2+c_s^2}\simeq v_{\rm w}\lesssim B_{\rm load}^2R_2^2/(\xi\dot M_{\rm w})$. For the adopted parameters, this formal limit is $1.3\times10^7\xi^{-1}(B_{\rm load}/100~{\rm G})^2~{\rm km\,s^{-1}}$, beyond the non-relativistic regime and indicating sufficient capture margin for stellar winds. At fixed mass supply and magnetic geometry, a higher launch speed lowers $\rho_{\rm w}\propto v_{\rm w}^{-1}$, while additional acceleration along the loaded tubes reduces $n_{\rm e}\propto v_\parallel^{-1}$ (Eq. \eqref{eq:tube_density}), making the local ECME density condition easier to satisfy.

\par
\begin{figure*}[htbp]
    \centering
    \includegraphics[width=\linewidth]{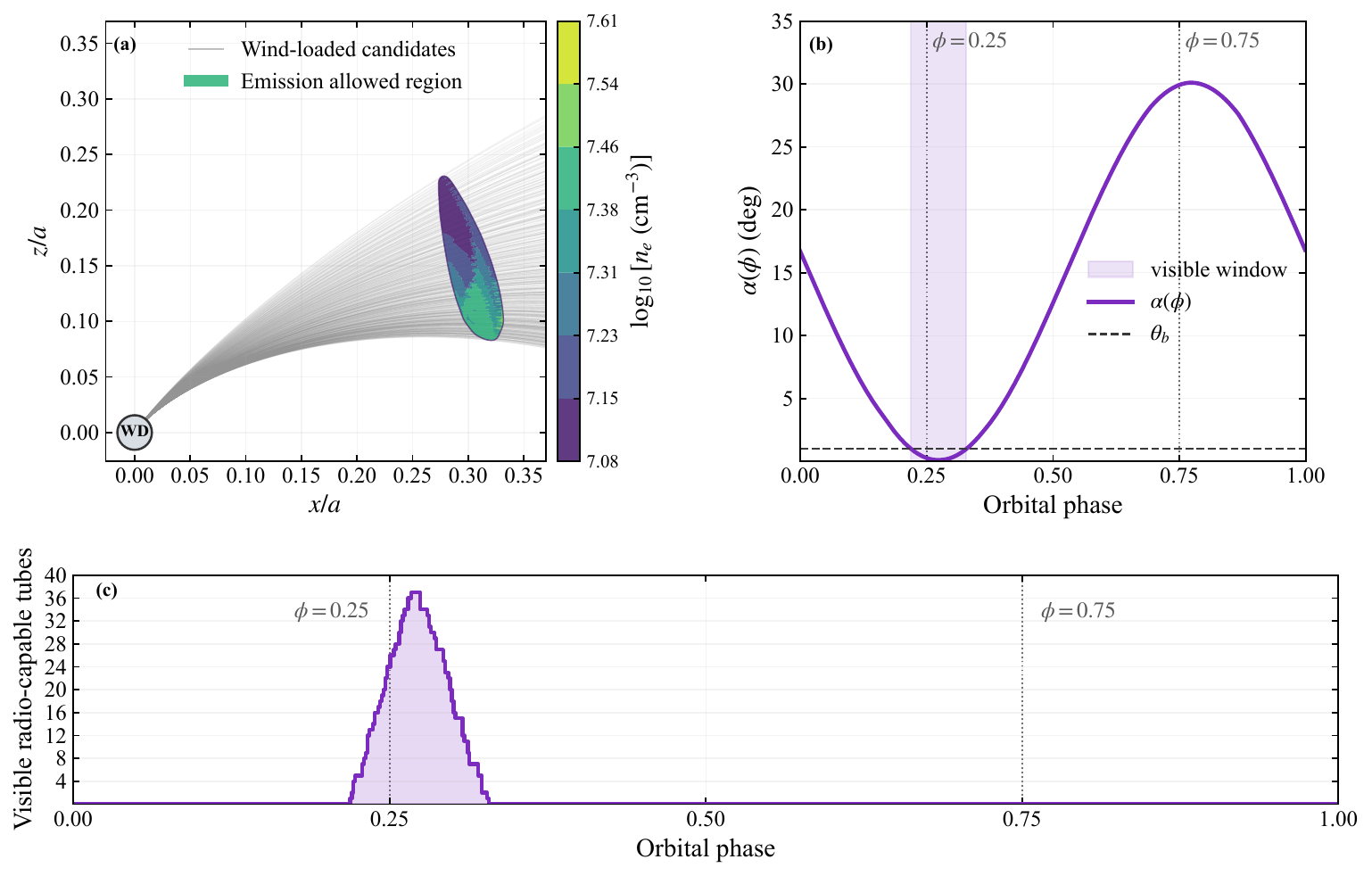}
    \caption{
        Radio-emission geometry and orbital visibility. (a) Projection of the wind-loaded candidate tubes onto the $x$--$z$ plane. The coloured contour, coded by electron density $\log_{10}(n_{\rm e}/{\rm cm^{-3}})$, demonstrates the region that satisfies all adopted cyclotron-frequency, maser-production, and beaming criteria at the reference phase. This coloured region is the envelope of accepted tubes: not every flux tube passing through its projected area is radio-capable.
        (b) Orbital variation of the emission--observer angular mismatch $\alpha(\phi)$ for a representative allowed sample. Visibility requires $\alpha(\phi)\leq\theta_b$. The shaded interval marks the resulting visibility window near the conjunction at $\phi=0.25$. 
        (c) Number of radio-capable flux tubes visible to the observer as a function of orbital phase. Visible tubes occur only within the window around $\phi=0.25$. Tube counts are a geometric visibility diagnostic and should not be interpreted as radio flux.
        }
    \label{fig:radio_profile}
\end{figure*}

 We evaluate the radio conditions at $\nu_\text{obs}=1 \text{ GHz}$, over a fractional bandwidth $\delta _\nu=0.05$. For the fundamental cyclotron harmonic emission ($s=1$), the frequency constrains the available magnetic field to $373–392 \text{ G}$. Intersecting this narrow field-strength layer with the accepted tubes confines the candidate source at a distance of approximately $0.3a$ from the WD. The coloured envelope in Fig. \ref{fig:radio_profile}(a) shows the part of field-strength layer that also satisfies the adopted maser and visibility conditions at the reference phase. 

\par Within the coloured region, the electron density spans $n_{\rm e}\sim(1-4)\times10^7~{\rm cm^{-3}}$, giving $\omega_{\rm p}/\omega_{\rm B}\sim0.03-0.06$, comfortably below the adopted ECME limit of 0.3. The entire emission region is compatible with the local low-density requirement.

\par The angular visibility of the radio emission relies on the emission angle. The reported ASKAP/ATCA median pulse phase is at $\phi=0.31\pm0.03$ \citep{rosePeriodicRadioXray2026}. An orbital-phase scan of the hollow-cone geometry shows that a family of nearby cone angles produces a single visibility interval around the conjunction at $\phi=0.25$ (see Append. \ref{app:scan} for the scan details). We adopt $\chi=69^\circ$ and $\theta_b=1^\circ$ as a representative hollow-cone geometry from this family. These angular parameters test whether the allowed field directions can place a visibility window within the observed ASKAP/ATCA phase range. 

\par For the adopted parameter, Fig. \ref{fig:radio_profile}(b) shows the variation of the emission-observer angular mismatch $\alpha(\phi)$ with respect to the orbital phase. The $\alpha(\phi)$ reaches its minimum near the conjunction $\phi=0.25$ and then increases smoothly as the binary rotates. One continuous geometric visibility window, $\phi=0.219-0.328$, is produced, under the visible condition $\alpha(\phi)<\theta_b$. The possible pulse width, $\Delta\phi=0.11$, corresponds to a maximum geometrical duty cycle of about $11\%$, before variations in the maser emissivity or propagation are considered.

\par Fig. \ref{fig:radio_profile}(c) shows the number of radio-capable flux tubes visible to the observer as a function of orbital phase. Under a total of $30\times 60$ samples, the number of visible tubes rises to a maximum of 37 near the center of the window and then returns to zero. The count measures geometric availability of emitting tubes and is not weighted by maser emissivity.

\par At the opposite conjunction, near $\phi=0.75$, the representative mismatch is about $30^\circ$, far larger than the adopted $1^\circ$ beam width, and none of the tubes in the emission region is visible. The fiducial solution therefore accounts for the primary radio window at $\phi\sim 0.25$ but not for radio events at the opposite conjunction. Possible changes in the active magnetic-tube family that could produce such events are considered in Sect. \ref{sec:discussion}.


\section{Discussion}\label{sec:discussion}

\subsection{X-ray energetics}\label{sec:xray_energetics}
\par For the adopted $M_1=0.83\,M_\odot$, $R_1=6.52\times10^8~{\rm cm}$, and $\dot M=5\times 10^{-11}\,M_\odot\,{\rm yr}^{-1}$, the gravitational power available at the WD surface is
\begin{equation}
    \begin{split}
     L_{\rm grav}&=\frac{GM_1\dot M}{R_1}\\
     &=5.36\times10^{32}~{\rm erg~s^{-1}}.
    \end{split}
\end{equation}
With the energy transfer efficiency $\eta_\text{X}=0.2$, the available accretion luminosity is 
\begin{equation}
    \begin{split}
        L_\text{acc}&=\eta_\text{X} L_{\rm grav}\\
        &=1.07 \times10^{32}~{\rm erg~s^{-1}}
     \label{eq:discussion_xray_gravitational_power}
    \end{split}
\end{equation}
which is consistent with the measured unabsorbed $0.5-10~{\rm keV}$ flux, $F_{\rm X}=(2.5\pm0.3)\times10^{-12}~{\rm erg~cm^{-2}~s^{-1}}$, giving \citep{imbrognoXrayEmissionLongperiod2026}
\begin{equation}
\begin{split}
 L_{\rm X,obs}
 &=4\pi d^2F_{\rm X}\\
 &=9.72\times10^{31} \left(\frac{d}{0.57~{\rm kpc}}\right)^2 {\rm erg~s^{-1}},
\label{eq:discussion_xray_observed_luminosity}
\end{split}
\end{equation}
The accretion power is thus comparable to the observed band luminosity.
\par The hard and soft X-ray temperatures characterize physically distinct regions and can constrain the model. 
For the loading radius $r_{\rm load}=0.13-0.28a=(5.6-11.9)\times10^9~{\rm cm}$, the calculated $h_{\rm sh}=0.01-0.11R_1$ limits give $kT_{\rm s,max}=36.2-38.8~{\rm keV}$, respectively, according to Eq. \eqref{eq:shock_temperature}. 

\par These values represent the upper edge of the post-shock temperature distribution, whereas the observed hard spectrum includes emission from gas cooling below the shock \citep{aizuXRayEmissionRegion1973,cropperEffectsGravityStructure1999}. In an approximately isobaric cooling flow, enthalpy conservation gives
\begin{equation}
 \frac{{\rm d}L}{{\rm d}T}=\frac{5k_{\rm B}\dot M}{2\mu_{\rm sh}m_{\rm p}}\simeq {\rm constant},
\label{eq:discussion_isobaric_cooling_luminosity}
\end{equation}
as adopted in the multi-temperature cooling-flow model in accreting white dwarfs \citep{mukaiTwoTypesXRay2003}. The luminosity-weighted temperature is then 
\begin{equation}
    \langle T\rangle_L=\frac{\int T(\mathrm d L/\mathrm d T)\mathrm d T}{\int (\mathrm d L/\mathrm d T)\mathrm d T}\simeq \frac{T_{\rm s,max}}{2}
\end{equation}
The resulting $k_B\langle T\rangle_L=18.1-19.4~{\rm keV}$, is compatible with the measured $k_BT_{\rm hot}=15^{+4}_{-3}~{\rm keV}$. The model prediction is therefore compatible with the observed hard-component temperature.

\par The soft component probes energy thermalized in the WD atmosphere. The fitted temperature, $k_{\rm B}T_{\rm cap}=0.14^{+0.03}_{-0.01}~{\rm keV}$, implies a large local surface flux. A $0.15~{\rm keV}$ blackbody requires $\sigma_{\rm B}T_{\rm cap}^4\simeq5\times10^{20}~{\rm erg~cm^{-2}~s^{-1}}$, where $\sigma_\text{B}$ is the Stefan--Boltzman constant. Direct thermalization relates the flux to the local accretion rate $\dot m_\text{loc    }$ through
\begin{equation}
 \sigma_{\rm B}T_{\rm cap}^{4}\simeq
 \epsilon_{\rm loc}\dot m_{\rm loc}GM_1
 \left(\frac{1}{R_1}-\frac{1}{r_{\rm load}}\right),
\label{eq:cap_thermalization_temperature}
\end{equation}
where $\epsilon_{\rm loc}$ is the fraction of the available energy thermalized locally. For the adopted parameters, $k_{\rm B}T_{\rm cap}=0.14$--$0.15~{\rm keV}$ requires $\epsilon_{\rm loc}\dot m_{\rm loc}\simeq(2.6$--$3.5)\times10^3~{\rm g~cm^{-2}~s^{-1}}$. In contrast, the local mass flux calculated by Eq. \eqref{eq:specific_mass_flux_lambda} is $\dot m_\text{sh}\sim 1 \,{\rm g\,cm^{-2}\,s^{-1}}$, which produces only $k_B T_{\rm cap}\sim 20~{\rm eV}$ for $\epsilon_\text{loc}=1$. Under the blackbody interpretation, the fitted soft temperature requires dense non-uniformity of the mass flux within the magnetic footprint. Dense blobs or other small-scale accretion inhomogeneities provide candidate mechanisms for this non-uniformity \citep{kuijpersCommentsRadialWhite1982,frankSoftXrayExcess1988}.

\par Irradiation and thermal conduction provide complementary routes for heating the footprint \citep{lambUVRadiationAccreting1979,imamuraXrayUltravioletRadiation1987}. For an isotropic source concentrated at the shock front, the irradiation temperature immediately beneath it is
\begin{equation}
 T_{\rm irr}=\left[
 \frac{(1-\mathcal A)L_{\rm irr}}
 {4\pi\sigma_{\rm B}h_{\rm sh}^2}
 \right]^{1/4},
\label{eq:discussion_irradiation_temperature}
\end{equation}
where $L_{\rm irr}$ is the irradiating luminosity, and $\mathcal A$ is the effective albedo. With $\mathcal A=0$, $L_{\rm irr}=10^{32}~{\rm erg~s^{-1}}$, and the calculated $h_{\rm sh}=0.01$--$0.11R_1$, this prescription gives $k_{\rm B}T_{\rm irr}\simeq0.006$--$0.02~{\rm keV}$. For the calculated shock temperature and shock height, a classical conductive estimate gives a downward heat flux of order, corresponding to approximately $10~{\rm eV}$ \citep{spitzerTransportPhenomenaCompletely1953}. Two-temperature calculations also find that bremsstrahlung cooling limits the power conducted to the WD surface \citep{imamuraXrayUltravioletRadiation1987,frankSoftXrayExcess1988}. 
Irradiation, thermalization, and thermal conduction are therefore limited to produce a hotter soft cap.

\par High-temperature soft components are also observed in soft IPs \citep{bernardiniBroadbandCharacteristicsSeven2017}, where compact, hotter regions within the accretion footprint have been proposed \citep{anzolinTwoNewIntermediate2008}. The inferred blackbody temperature depends on the treatment of ionized absorption and emission lines \citep{islamRoleComplexIonized2021}. The emitting structure and spectral decomposition are therefore both uncertain, and we adopt the measured soft temperature as an empirical input without resolving its detailed heating mechanism.

\subsection{Radio energetics}\label{sec:discuss_radio_energetics}

\par For the central distance $d=0.57~{\rm kpc}$, a representative $F_\nu=5~{\rm mJy}$ pulse at $\nu=1~{\rm GHz}$ has, within the $\delta_\nu=0.05$ band used in our
calculation, the isotropic luminosity is estimated by \citep{rosePeriodicRadioXray2026}
\begin{equation}
\begin{split}
 L_{\rm R,iso}&=4\pi d^2F_\nu\Delta\nu\\
 &=9.72\times10^{25}\left(\frac{d}{0.57~{\rm kpc}}\right)^2\left(\frac{F_\nu}{1~{\rm mJy}}\right)\\
 &\quad\times\left(\frac{\nu}{5~{\rm GHz}}\right)\left(\frac{\delta_\nu}{0.05}\right){\rm erg~s^{-1}}.
 \label{eq:radio_isotropic_luminosity}
\end{split}
\end{equation}

Since the radio emission is highly beamed in the hollow annulus between $\chi-\theta_b$ and $\chi+\theta_b$, the radio energy budget should be corrected by the beaming angle fraction $f_b=\Omega_b/4\pi\simeq \sin\chi \sin \theta_b\sim 1.6\times 10^{-2}$. The corresponding radiated power is therefore $L_{\rm R}=f_b L_{\rm R,iso}\simeq 1.56\times10^{24}~{\rm erg~s^{-1}}$. 

\par The gravitational energy release is a natural reservoir for the radio-emission energy. The representive loading and emission radii are $r_\text{load}=a-R_2\simeq 0.78a$ and $r_\text{emission}\simeq 0.3a$, respectively. The gravitational potential drop along the loaded tubes is 
\begin{equation}
    \begin{split}
     \dot E_{\rm grav,R}&= GM_1\dot M_\text{w,vis}\left(\frac{1}{r_{{\rm emission}}}-\frac{1}{r_{{\rm load}}}\right)\\
     &\simeq 4.48\times10^{24}~{\rm erg~s^{-1}}.
    \label{eq:wind_gravity}
\end{split}
\end{equation}
where $\dot M_\text{w,vis}=f_\Omega \dot M_\text{w}$ is the mass-loss rate along the visible tubes weighted by their solid angular fraction $f_\Omega\sim 10^{-3}$ with respect to the whole MD. The radio energy budget for the gravitational potential drop is therefore
\begin{equation}
\begin{split}
 L_{\rm R,grav}&=\eta_{\rm R,grav}\dot E_{\rm grav,R}\\
 &\simeq 4.48\times10^{23}\left(\frac{\eta_{\rm R,grav}}{0.1}\right)~{\rm erg~s^{-1}}.
 \label{eq:wind_gravity_radio_power}
\end{split}
\end{equation}
where $\eta_\text{R,grav}$ is the effective transfer factor. The $L_\text{R,grav}$ can barely satisfy the energy budget when $\eta_\text{R,grav}\gtrsim 0.36$. Thus, local gravitational release is a possible but very restrictive power source.


\par The fiducial model above adopts a synchronous, time-independent geometry, where the WD dipole is fixed in the corotating frame. This simplifying assumption isolates how the joint field maps the dense stream and tenuous wind onto distinct flux tubes and determines their visibility. It is a modeling choice rather than an independent measurement of the WD spin. However, an asynchronous binary with differential rotation between the WD magnetosphere and the orbit then provides an additional driver for particle acceleration, which should ease the energy budget problem. We define $\delta_\Omega\equiv|\Omega_{\rm s}-\Omega_\text{orb}|/\Omega_\text{orb}$, where $\Omega_\text{s}$ is the WD spin. At the donor, $B_{\rm WD}(a)=\mu_1/a^3=20.4~{\rm G}$, compared with $B_2=100~{\rm G}$. The field comparison indicates that the WD and MD magnetospheres should interact directly, leading to magnetic reconnection events that accelerate particles \citep{yangMagneticWhiteDwarf2026}.
\par For the magnetic-interaction prescription, the large-scale interacting-dipole torque $N_B\sim\mu_1\mu_2/a^3$ gives
\begin{equation}
\begin{split}
    \dot E_{\rm MI}&\sim N_B|\Omega_{\rm s}-\Omega_\text{orb}|\sim\frac{\mu_1\mu_2\Omega_\text{orb}}{a^3}\delta_\Omega\\
    &\simeq 1.95\times10^{30}\delta_\Omega~{\rm erg~s^{-1}}.
\end{split}
 \label{eq:MI_power}
\end{equation}
Given $\eta_{\rm R,MI}$ as the net fraction of the magnetic-interaction power to the radio emission in the modeled band, the available energy budget should be $L_\text{R,MI}=\eta_{\rm R,MI}\dot E_{\rm MI}$, which is fair enough to explain the radio emission even for a near-synchronous binary. Combining $L_\text{R,MI}\geq f_bL_{\rm R,iso}$ with the beat relation $P_{\rm beat}={2\pi}/{|\Omega_{\rm s}-\Omega|}=P_{\rm orb}/\delta_\Omega$ gives
\begin{equation}
 P_{\rm beat}^{\rm MI}\lesssim 180~{\rm d} \left(\frac{\eta_{\rm R,MI}}{0.25\%}\right).
 \label{eq:beat_period_limit}
\end{equation}
Thus, for a conservative choice of $\eta_{\rm R,MI}$, a month-scale beat period is allowed. An independent spin measurement or repeated monitoring is required to measure the actual beat clock.

\subsection{Radio phase switching}\label{sec:phase_switching}

\par J1745 shows epoch-dependent radio emission at $\phi=0.25$ and $\phi=0.75$. We interpret the epoch-dependent radio phases using quasi-static magnetic configurations at different beat phases. The steady calculations in this work should be regarded as quasi-static snapshots at specified beat phases. As the WD orientation changes relative to the donor, different connected tube families can become visible. 

\begin{figure*}[htbp]
    \centering
    \includegraphics[width=\linewidth]{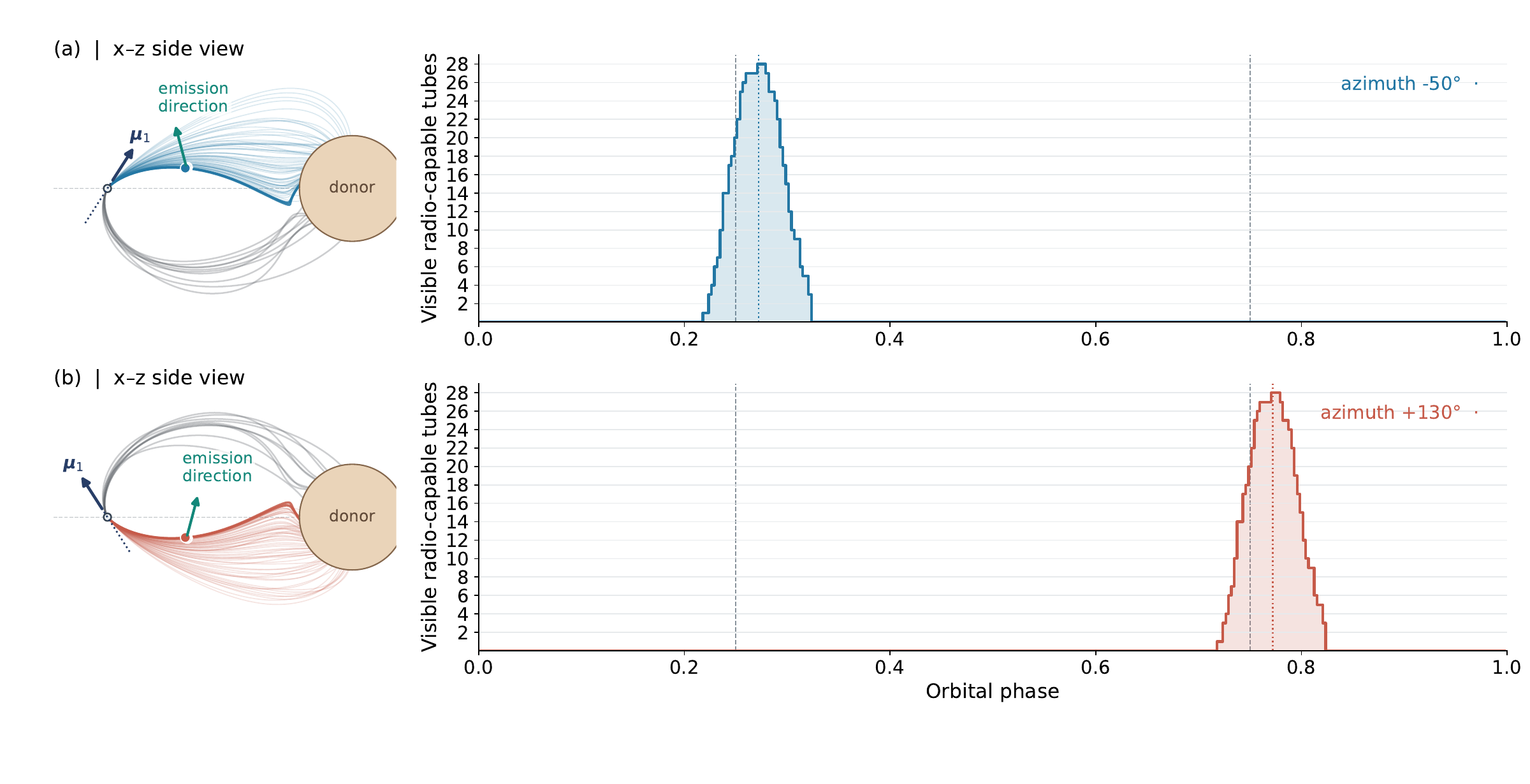}
    \caption{Two quasi-static radio geometries illustrating an epoch-dependent phase solution. Row (a) shows the configuration with WD magnetic azimuth $\psi_{\rm WD}=-50^\circ$ and wind loading onto the northern pole, while row B shows the configuration with $\psi_{\rm WD}=+130^\circ$ and wind loading onto the southern pole. The left panels show projections of the three-dimensional field lines onto the $x$–$z$ plane. Blue and red curves represent wind-loaded flux tubes connected to the selected pole in rows (a) and (b), respectively, with radio-capable tubes shown as thicker lines. Gray curves represent field lines connected to the opposite pole. Colored points mark candidate radio-emission regions, and the green arrows schematically indicate the projected emission directions toward the observer. The right panels count the number of radio-capable flux tubes visible to the observer as a function of orbital phase. The visibility windows span orbital phases $0.22-0.33$ for $\psi_{\rm WD}=-50^\circ$ and $0.73-0.82$ for $\psi_{\rm WD}=+130^\circ$.}
    \label{fig:epoch_radio_two_pole}
\end{figure*}

\par To connect wind-loading geometry to a possible beat-cycle evolution, we adopt a geometric pole-dominance prescription in which donor-wind material is preferentially loaded to field lines associated with the magnetic pole oriented more closely toward the donor. Similar pole-switching behavior has been observed or simulated in near-synchronous polars \citep{ramsayRXJ21155840Confirmation1999,hakalaTESSObservationsAsynchronous2019}, where beat cycle redirects accretion between magnetic poles. We adopt this prescription only as a geometric criterion for selecting the preferred magnetic hemisphere, rather than as a self-consistent calculation for the capture efficiencies of the two poles. In the present geometry, with the donor on the $+x$ axis and the WD dipole tilted by $45^{\circ}$, we assume that the wind accretion is dominated by the northern pole over $-90^\circ<\psi_{\rm WD}<+90^\circ$ (the $+x$ hemisphere), while it is dominated by the southern pole over $+90^\circ<\psi_{\rm WD}\leq+180^\circ$ and $-180^\circ\leq\psi_{\rm WD}<-90^\circ$ (the $-x$ hemisphere).

\par Under the prescription, we find an explicit two-state solution for the epoch-dependent radio phase within the adopted wind-loading model. The two calculations retain the same viewing inclination, wind prescription, $1$-GHz band, and the ECME emission angle $\chi=69^\circ$ with $\theta_b=1^\circ$. Fig~\ref{fig:epoch_radio_two_pole} compares the radio visibility of two quasi-static WD orientations. At WD magnetic azimuth $\psi_{\rm WD}=-50^\circ$, the wind accretion onto the north pole is dominated due to the magnetic pole orientation. This case produces a radio-visible window at orbital phase $\phi=0.22-0.33$ with its maximum at $\phi=0.27$. By contrast, at $\psi_{\rm WD}=+130^\circ$, wind-loading tubes ending on the southern pole produce a visible window at 0.73-0.82 with a maximum at $\phi=0.769$. The resulting phases lie near the ASKAP/ATCA and MeerKAT burst groups, whose measured median phases are $0.31\pm0.03$ and $0.8\pm0.1$, respectively \citep{rosePeriodicRadioXray2026}. Thus, this scenario supplies a concrete geometrical solution for both observed phase windows.

\par We emphasize that, near $\psi_{\rm WD}=\pm 90^\circ$, the two poles are equally close to the donor under this geometric criterion. If the donor wind simultaneously loads into the flux-tube families connected to both poles, and both satisfy the ECME conditions, two radio outbursts could be seen within one orbital cycle. For example, a short interval of two-pole accretion has been observed in the asynchronous polar CD Ind \citep{hakalaTESSObservationsAsynchronous2019}. Alternatively, the radio emission may temporarily shut down during this period if the transition prevents flux-tube families from satisfying the ECME condition due to the magnetic and loading configuration change. Our current pole-switching prescription is a simplified explanation of epoch-dependent radio geometry. The scenario would be further supported if long-term monitoring between the two observed radio epochs revealed either a two-burst pattern or a temporary radio shutdown.

\par The beat interpretation can be illustrated by a possible timing solution for the observing epochs. The ATCA observations were obtained in several campaigns spanning days to weeks, with some successive campaigns separated by roughly half a year, whereas the three MeerKAT epochs were concentrated within one week
\citep{rosePeriodicRadioXray2026}.  The late ATCA observation (Epoch 8) began on 2024 September 15, with an observation of a radio burst at $\phi=0.25$. The first MeerKAT observation began on 2024 December 27 with a burst at $\phi=0.75$ \citep{rosePeriodicRadioXray2026}.  Two observations are separated by $103.10~{\rm d}$. For an illustrative beat period $P_{\rm beat}=180~{\rm d}$, the separation corresponds to a beat phase difference of $\Delta\Phi_{\rm beat}=0.573$,  or approximately half a beat cycle.  A 180-d beat clock is therefore compatible with the ASKAP/ATCA and MeerKAT observations sampling different beat phases.  This prescription predicts recurrence of the same radio state after approximately 180 d and transitions toward the other state on a roughly 90-d separation. Long-term monitoring over several such cycles can test the beat prescription directly.

\par The beat should change the global threading geometry. The dense stream should therefore be captured onto a different footprint, and the phase or shape of the X-ray modulation should vary coherently on the same beat clock. If the beat exists, X-ray phase shift or long-timescale modulation should be observed (illustrated in Append. \ref{app:xray_modulation}). Simultaneous radio and X-ray observations near the two radio states would be particularly valuable for identifying the loading model.

\subsection{J1745 and the evolutionary LPT--mCV connection}
\par Coexistence of magnetic accretion and LPT-like pulses in J1745 raises the possibility that periodic radio activity can occur at more than one stage of mCV evolution \citep{rosePeriodicRadioXray2026,kniggeLongPeriodRadioTransient2026}. In our framework, persistent pulses require a connected low-density channel and a source of energetic electrons. We examine how these conditions differ between ordinary polars, IPs, and a low-rate system near the orbital-period minimum.

\subsubsection{Obstacles to LPT-like emission in ordinary mCVs}\label{sec:ordinary_mcv_non_lpt}
\par Ordinary mCVs do show polarized radio flares, but not stably periodic LPT-like radio emission \citep{barrettRadioObservationsMagnetic2020}. In our model, such emission in mCVs requires a clean, WD-connected, low-density loading channel with an electron accelerator.  Polars and IPs can encounter different obstacles to satisfying these requirements.

\par In a strictly synchronized polar, the magnetic interaction due to the spin--orbit asynchronicity vanishes, $\dot E_{\rm MI}=|N_B(\Omega_{\rm s}-\Omega_{\rm orb})|=0$ (Eq.~\eqref{eq:MI_power}). Synchronization therefore removes this additional source of electron acceleration for the radio emission. Donor stellar activity can provide plasma acceleration and electron injection, but it may not provide a secularly stable coherent-radio-emission environment. Gravitational acceleration remains available, but the energy budget is limited (see Sect. \ref{sec:discuss_radio_energetics}). Converting this limited energy reservoir into radio emission still requires a mechanism capable of energizing the electrons and establishing the non-thermal, anisotropic velocity distribution required for coherent maser emission \citep{treumannElectronCyclotronMaser2006}. 

\par Strong fields of polar also bring part of the accretion curtain close to the wind branch. Considering a polar with $P_\text{orb}=2$h, $M_1=0.8\,M_\odot$, $M_2=0.2\,M_\odot$, $B_{1}=30$ MG, and $\dot M=10^{-10}\,M_\odot\,{\rm yr^{-1}}$, representative of observed short-period polars \citep{beuermannNeglectedXrayDiscovered2021}, the ballistic stream is captured at the L1 point, with curtain boundary $q=0.01$. The high magnetosphere loading for the curtain indicates potential pollution of the wind-loaded tubes. The WD footprints of the L1-loaded stream and those of the GHz-capable wind tubes are only $\sim 0.13^\circ$ apart, corresponding to $\sim 14$ kilometers. 

\par A very minor adjacent matter leak can pollute the radio-capable tubes. For our permissive $\omega_{\rm p}/\omega_{\rm ECME}<0.3$ criterion, the local plasma density limitation is
\citep{treumannElectronCyclotronMaser2006,willesElectroncyclotronMaserEmission2004}
\begin{equation}
 \begin{split}
 n_{\rm crit}(\nu)&=\frac{m_{\rm e}}{4\pi e^2}(0.3\times 2\pi\nu)^2\\
 &\simeq1.12\times10^9
 \left(\frac{\nu}{1~{\rm GHz}}\right)^2~{\rm cm^{-3}}.
 \end{split}
 \label{eq:discussion_radio_density_threshold}
\end{equation}
Using Eq.~\eqref{eq:tube_density} with $\Lambda_\text{w}\sim 10^{-11}{\rm g~ cm^{-2}~s^{-1}~G^{-1}}$, the plasma density of the wind-loaded tubes in GHz region is $n_e\sim 10^7\text{ cm}^{-3}<n_{\rm crit}(\nu)$. If the accretion flow leaks mass with $\simeq 3.2\times10^{-13}\,M_\odot\,{\rm yr^{-1}}$, or $0.32\%$ of the stream into the wind-loaded tubes, the injection is enough for the wind-loaded tubes to reach the local plasma density limitation. This susceptibility may prevent the polar from retaining stable, clean wind-loaded tubes for coherent radio emission.

\par In longer-period IPs, the difficulty can instead be maintaining a stable WD-connected wind channel. For an illustrative 4-h IP with $M_2=0.5\,M_\odot$, $R_2\simeq0.5\,R_\odot$, and $B_{1}=5.4$ MG, retaining $M_1$ and the wind prescription above, the WD contribution is only $\lesssim 4$ G near the donor, well below the adopted $\sim 100$ G donor surface field. The wind-launching region is therefore MD-dominated. The plasma is initially guided primarily by the donor field, rather than automatically loaded onto WD-connected tubes. Plasma on loops that close back onto the MD can be confined, whereas escaping wind follows open donor flux \citep{vidottoMdwarfStellarWinds2014}. Feeding a WD-connected channel thus requires either capture of escaping wind or a magnetic connection between the stars. The efficiency of this transfer depends sensitively on field geometry and orbital separation \citep{cohenROLEWHITEDWARFS2012}. MD dominance does not prevent wind accretion, but makes sustained feeding of a particular low-$\Lambda$ channel an additional topological requirement. 

\par Disk-fed IPs introduce a further complication to the magnetic connectivity. The magnetospheric truncation and corotation radii are\citep{ghoshAccretionRotatingMagnetic1979,frankAccretionPowerAstrophysics2002}
\begin{equation}
 r_{\rm m}=k_{\rm m}\left(\frac{\mu_1^4}{2GM_1\dot M^2}\right)^{1/7},
 \qquad
 r_{\rm co}=\left(\frac{GM_1P_{\rm s}^2}{4\pi^2}\right)^{1/3},
 \label{eq:discussion_magnetosphere_corotation}
\end{equation}
where $P_{\rm s}$ is the WD spin period and $k_{\rm m}=0.5$ accounts approximately for the disk geometry. The illustrative 4-h IP example gives $r_{\rm m}\simeq0.086a$ and $r_{\rm co}\simeq0.144a$ for an assumed $P_{\rm s}=1000$ s. The accretion-stream circularization radius is $r_{\rm circ}\simeq 0.149a$ \citep{frankAccretionPowerAstrophysics2002}. Once a disk is present, differential rotation between the WD and the disk can twist, inflate, and open magnetically linked flux, forming current sheets \citep{uzdenskyPartialFieldOpening2002}. Related IP simulations with a disk produce an extended hot corona, deform the stellar magnetosphere, and predict thermal emission in the 0.1--10 keV band \citep{barberaAccretionDiskCoronae2017}. These results demonstrate that disk-associated activity can modify the magnetic and plasma environment, although they do not establish coherent radio suppression. In our model, such environmental modification may disrupt WD--MD connectivity or enable plasma from the disk to leak into the low-$\Lambda$ wind channel due to disk reconnection. We therefore regard the connectivity disruption or contamination as a possible obstacle to persistent LPT-like emission, rather than an inevitable consequence of disk formation. 

\par These examples indicate that persistent coherent radio emission requires a suitable magnetic field, geometry, and mass supply. Strong fields can load a stream close to wind-connected flux, while donor field-line closure, disk activity, or increased upstream contamination may suppress a persistent low-$\Lambda$ channel. A lower total $\dot M$ is favourable only if the channel remains connected and supplied with accelerated electrons. These conditions allow a possible LPT-like activity in suitable evolutionary stages.

\subsubsection{Evolutionary ordering and the status of J1745}
\par A detached magnetic WD-MD binary can form a LPT system before Roche-lobe overflow. After contact, it becomes an mCV. It can therefore be called ``pre-mCV'' in the detached state \citep{yangMagneticWhiteDwarf2026}. The relevant evolutionary sequence is a detached pre-mCV, an accreting mCV, and a period bouncer \citep{breedtEvolutionaryStateShortperiod2012,pretoriusSpaceDensityMagnetic2013,yangMagneticWhiteDwarf2026}.
 

\par We analyse the evolution, primarily dominated by gravitational dissipation and mass transfer. For a circular binary, take conservative Roche-lobe transfer with $\dot M=-\dot M_2>0$, neglecting spin angular momentum and the direct mass-loss terms of a weak escaping wind. With $J_{\rm orb}=M_1M_2[G a/M]^{1/2}$, the mass-transfer (MT) contribution to the orbital evolution is \citep{frankAccretionPowerAstrophysics2002}
\begin{equation}
 \left(\frac{\dot P_{\rm orb}}{P_{\rm orb}}\right)_{\rm MT}=3\frac{\dot M}{M_2}\left(1-\frac{M_2}{M_1}\right).
 \label{eq:discussion_period_mt}
\end{equation}
\par Gravitational-wave (GW) dissipation contributes \citep{petersGravitationalRadiationMotion1964}
\begin{equation}
 \left(\frac{\dot P_{\rm orb}}{P_{\rm orb}}\right)_{\rm GW}
 =-\frac{96}{5}\frac{G^{5/3}}{c^5}\frac{M_1M_2}{M^{1/3}}\Omega_\text{orb}^{8/3}.
 \label{eq:discussion_period_gw}
\end{equation}
Thus, the total orbital evolution is the sum of both contributions
\begin{equation}
\begin{aligned}
 \frac{\dot P_{\rm orb}}{P_{\rm orb}}
 &=\left(\frac{\dot P_{\rm orb}}{P_{\rm orb}}\right)_{\rm MT}
  +\left(\frac{\dot P_{\rm orb}}{P_{\rm orb}}\right)_{\rm GW}.
\end{aligned}
 \label{eq:discussion_period_competition}
\end{equation}
The sign of $\dot P$ indicates that MT widens the orbit whereas GW shrink it, for $M_2<M_1$.

\par At the detached stage, with $P_\text{orb}\gtrsim 2\text{ h}$, the orbital evolution of a detached pre-mCV is dominated by GW dissipation, resulting in $\dot P_\text{orb}<0$. The orbit therefore continues to shrink until the donor fills its Roche lobe. For typical parameters with $M_1=0.8\,M_\odot$, $M_2=0.2\,M_\odot$ and $P_\text{orb}=2\, \text{h}$, the gravitational dissipation gives $\dot P_{\rm orb}=-2\times 10^{-13} \text{ s s}^{-1}$. The binary remains detached on the timescale of $|P_\text{orb}/\dot P_\text{orb}|\sim 1 \text{ Gyr}$. In this pre-mCV stage, mass transfer is negligible, which provides a clean environment for the LPT-like emission. 


\par Once the donor fills its Roche lobe, sustained mass transfer begins, and the binary enters the accreting mCV phase.  Roche-lobe overflow initiates sustained mass transfer and the binary enters the accreting mCV phase. Observationally, the majority of mCVs in this stage are characterized by accretion-powered X-ray emission. At the same time, the dense accretion flow in the magnetosphere is expected to suppress the low-density conditions favorable for LPT-like coherent radio emission, as we discussed in the previous section.

\par The orbit keeps evolving under the MT and GW. At low mass ratio, the Roche-lobe contact implies 
\citep{kniggeEVOLUTIONCATACLYSMICVARIABLES2011}
\begin{equation}
 \frac{\dot P_{\rm orb}}{P_{\rm orb}}
 \simeq\frac{3}{2}\frac{\dot M}{M_2}\left(\frac{1}{3}-\zeta_2\right)
 \label{eq:discussion_period_donor_response}
\end{equation}
where where $\zeta_2={\mathrm{d}\ln R_2}/{\mathrm{d}\ln M_2}$. The orbit keeps shrinking for $\zeta_2>1/3$, where the GW dissipation dominates, and reaches its minimum at $1/3$. It widens after bounce for $\zeta_2<1/3$, where MT dominates. The MT dominance can be reached by the decrease of $M_2$, while the increase in $\dot M$ is unnecessary. The transition point is called the period bouncers, with a typical period of 1.3-1.7 h \citep{inightCataclysmicVariablesSloan2023}. After the period bouncer, the orbit starts expanding, though $M_2$ and $\dot M$ decrease. 

\par We place J1745 near this period bouncer. Its period $P_{\rm orb}\simeq 82$ min and donor mass $\simeq0.05-0.1\,M_\odot$ lie close to the empirical minimum, which favors a late, near-bouncer stage of the adopted donor sequence \citep{rosePeriodicRadioXray2026,kniggeLongPeriodRadioTransient2026}. The sequence implies $\dot M\simeq 10^{-11}\,M_\odot\,{\rm yr^{-1}}$ at the bouncer \citep{kniggeEVOLUTIONCATACLYSMICVARIABLES2011}, consistent with the $\dot M\simeq8\times10^{-11}\,M_\odot\,{\rm yr^{-1}}$ from J1745's observation \citep{kniggeLongPeriodRadioTransient2026} and our adopted $\dot M=5\times10^{-11}\,M_\odot\,{\rm yr^{-1}}$. The period, donor mass, and accretion-rate consistency support J745 as a low-rate, evolved mCV and a period-bouncer candidate. 

\par This placement indicates a potential evolutionary history for J1745-like systems. In a detached pre-mCV, a tenuous donor wind can supply WD-connected tubes, while asynchronous magnetic interaction can provide an electron accelerator \citep{laiDCCIRCUITPOWERED2012,yangMagneticWhiteDwarf2026}. The orbit shrinks to contact due to gravitational and electromagnetic dissipation.
After Roche-lobe overflow begins, a large accretion rate and the resulting stream or disk can suppress the separate wind channel through the processes discussed in Sect.~\ref{sec:ordinary_mcv_non_lpt}. The system may then lose persistent LPT-like pulses during its ordinary accreting mCV phase. 

\par As the system approaches the period minimum, declining mass transfer may reopen a radio-capable channel if the wind-fed tubes remain separated from the Roche-lobe overflow stream and a source of energetic electrons persists. Our fiducial J1745 solution supplies a possible geometry for this stage: the Roche-lobe overflow feeds a high-$\Lambda$ curtain close to the WD, while the stellar wind is captured in the high magnetosphere as low-$\Lambda$ flux tubes.
The fiducial J1745 geometry illustrates this late-stage possibility: Roche-lobe overflow supplies a dense near-WD curtain, while a donor wind feeds separate low-loading tubes. We propose an evolutionary sequence in which radio activity is accessible before contact, can be suppressed during stronger accretion, and can reappear in suitable low-rate configurations. 

\par Establishing the proposed evolutionary connection between LPTs and mCVs requires larger, well-characterized samples and continuous multiwavelength observations. With sufficient observational coverage, combining radio properties with orbital periods, donor masses, and accretion rates could reveal how LPTs relate to different mCV populations and constrain the evolutionary pathways connecting detached pre-mCVs, accreting mCVs, and period bouncers \citep{kniggeEVOLUTIONCATACLYSMICVARIABLES2011}.

\section{Conclusion}\label{sec:conclusion}

\par J1745 connects two observational phenomena, accretion-powered X-rays and coherent, periodically visible radio emission. Its identification as an accreting magnetic WD binary extends the binary interpretation of LPTs into the mCV population. LPT-like emission is rarely observed in ordinary mCVs. The central physical question is therefore how a magnetosphere can load a dense accretion flow while maintaining sufficiently tenuous regions for coherent radio emission.

\par We have investigated this question with a unified mass-loading framework in the combined magnetic field of the WD and its donor. We trace a dense ballistic stream from Roche-lobe overflow and a stellar wind from the donor to their respective first-loading locations and then along WD-connected field lines. The conserved mass flux per unit magnetic flux, $\Lambda=\mathrm{d}\dot M/\mathrm{d}\Phi$, determines the density along both populations through $n_{\rm e}=\Lambda B/(\mu_{\rm e}m_{\rm p}v_\parallel)$. This construction links the candidate X-ray and radio regions to their plasma supply within a common binary geometry, without assigning an independent density to the radio source.

\par Our main conclusions are as follows.
\begin{enumerate}
\item In the high-$\Lambda$ branch, Roche-lobe overflow supplies a dense accretion curtain, a post-shock hard X-ray source, and a soft X-ray footprint near the WD. For the adopted stream rate of $5\times10^{-11}\,M_\odot\,{\rm yr^{-1}}$, the calculated loading rates are $\sim 10^{-6}~{\rm g\,cm^{-2}\,s^{-1}\,G^{-1}}$ in the loaded tubes. The calculated footprint determines the X-ray maximum near $\phi\simeq0.886$ by its orientation. The geometrical modulation of the X-ray sources and curtain absorption account for the X-ray variability. The calculated pulse fraction is $28.00\%$, $16.15\%$, and $20.06\%$ in the soft, hard, and full bands, respectively, which is consistent with the observation.

\item In the low-$\Lambda$ scenario, a donor wind with $\dot M_{\rm w}=10^{-14}\,M_\odot\,{\rm yr^{-1}}$ loads separate WD-connected tubes. The resulting loading rates are $\Lambda\simeq(6$--$23)\times10^{-12}~{\rm g\,cm^{-2}\,s^{-1}\,G^{-1}}$. Within GHz-emitting candidate regions in the tubes, the electron density spans $n_{\rm e}\sim(1-4)\times10^7~{\rm cm^{-3}}$, giving $\omega_{\rm p}/\omega_{\rm ECME}\sim0.03-0.06$, satisfying the adopted local ECME density criterion. With the representative hollow-cone geometry, the visible interval spans $\phi\simeq0.22-0.33$, compatible with  the primary ASKAP/ATCA pulse window. 

\item A small spin--orbit asynchronicity provides additional energy budget and a link for the observed changes in radio-pulse phase. Magnetic interaction can supplement the limited gravitational energy budget. The beat-dependent magnetic configuration can shift the radio visibility window. Under the adopted pole-dominance prescription, two quasi-static configurations produce radio-visibility maxima near $\phi\simeq0.27$ and $0.77$, respectively, providing a possible geometric explanation for the two epoch-dependent pulse windows. An illustrative $P_{\rm beat}\simeq180~{\rm d}$ could associate the two states with different beat phases, although the present calculations neither constrain this period nor follow a continuous switching trajectory. Changes in the global threading geometry may also shift the accretion footprint and alter the phase or shape of the X-ray modulation, providing an observational test of this interpretation.

\item These results suggest a possible evolutionary connection from detached pre-mCVs through accreting mCVs to period bouncers. Before Roche-lobe contact, a weak donor wind can supply a clean low-density radio channel. During the accreting mCV stage, stream contamination, changes in magnetic connectivity, or the absence of the electron acceleration driver may hinder persistent LPT-like emission. Near the period bouncer, declining mass transfer may allow a distinct low-$\Lambda$ channel to reproduce LPT-like emission. J1745's approximately $82$-min orbital period, low donor mass, and low accretion rate are consistent with such a late-stage configuration. The scenario provides a potential evolutionary history for J1745-like systems and could be tested through population studies.
\end{enumerate}

\par The present model provides a physical framework for the coexistence of a dense accretion curtain for X-ray emission and low-density radio flux for radio emission in J1745. It supports the possibility that J1745 is an evolved mCV linking LPT-like radio emission to a late stage of mCV evolution.

\section*{Acknowledgments}
This work was supported by the National Natural Science Foundation of China (grant Nos. 12625304 and 12494574).

\appendix
\section{Orbital-phase scan of the hollow-cone geometry}\label{app:scan}
\par
\begin{figure*}[htpb]
    \centering
    \includegraphics[width=\linewidth]{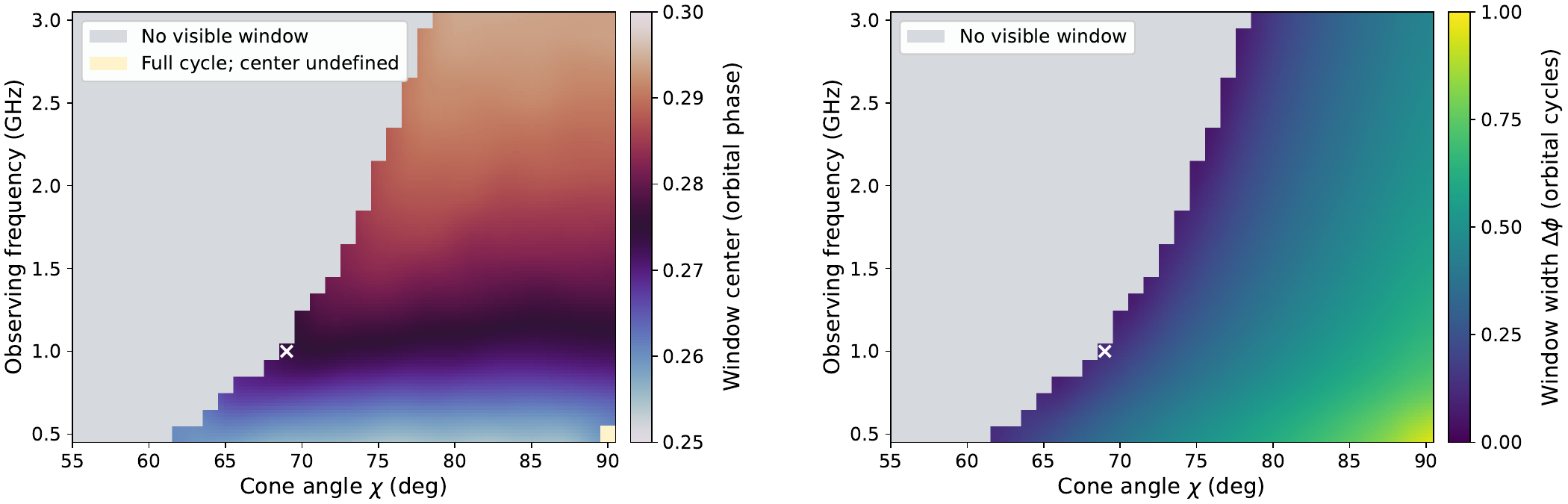}
    \caption{Sensitivity of the orbital radio-visibility windows to the ECME emission angle $\chi$ and observing frequency $\nu_{\rm obs}$. The left panel gives the central phase of the visibility window, and the right panel gives its width $\Delta\phi$ in orbital cycles. Gray denotes parameter combinations with no visible window, while pale yellow in the left panel indicates full-cycle visibility, for which the window center is undefined. White crosses mark the fiducial parameter combination in the main context. The colors describe geometric visibility, rather than radio flux or an emissivity-weighted pulse profile.}
    \label{fig:chi_freq_scan}
\end{figure*}

 We examine the dependence of orbital radio visibility on the ECME emission angle $\chi$ and observing frequency $\nu_{\rm obs}$. Fig. \ref{fig:chi_freq_scan} shows the visibility-window centers and widths over $\chi=55^\circ-90^\circ$ and $\nu_{\rm obs}=0.5-3~{\rm GHz}$, using a wind-source grid of $N_\alpha\times N_\psi=18\times36$. The binary parameters, joint magnetic field, and wind-loading prescription are fixed at their fiducial values. At each parameter pair, the candidate emission regions and the local plasma, ECME, escape, and beaming conditions are re-evaluated. A window center is the midpoint of a connected visible phase interval, accounting for orbital periodicity; it is undefined for full-cycle visibility. White crosses mark the fiducial parameter combination in the main context.

In Fig. \ref{fig:chi_freq_scan}, gray denotes parameter combinations with no visible window, while pale yellow in the left panel indicates full-cycle visibility, for which the window center is undefined. Across the visible region, window centers remain near $\phi\simeq0.25-0.30$. The widths depend more strongly on the scanned parameters: increasing $\chi$ generally broadens the window, whereas increasing frequency at fixed $\chi$ narrows it and can suppress visibility.

\par This scan supports the viability of our fiducial parameters, which provide one representative solution within an extended family of allowed radio geometries. Across most of the radio-visible parameter domain, the windows lie in the same observed phase interval around 0.25. Parameter combinations near the visibility boundary can yield windows with $\Delta\phi\sim0.1$, with comparable widths obtained at different frequencies by adjusting $\chi$. Therefore, the calculated pulse phase and available duration can be accommodated over a broad range of $\chi$ and $\nu_{\rm obs}$, without requiring a uniquely tuned parameter pair. This broad allowed region provides flexibility when selecting the emission angle for a given observing frequency. The widths describe the time available for visibility and do not specify the maser emissivity or burst duration.

\section{The magnetosphere-dependent X-ray modulation}\label{app:xray_modulation}
\par
\begin{figure*}[htbp]
    \centering
    \includegraphics[width=\linewidth]{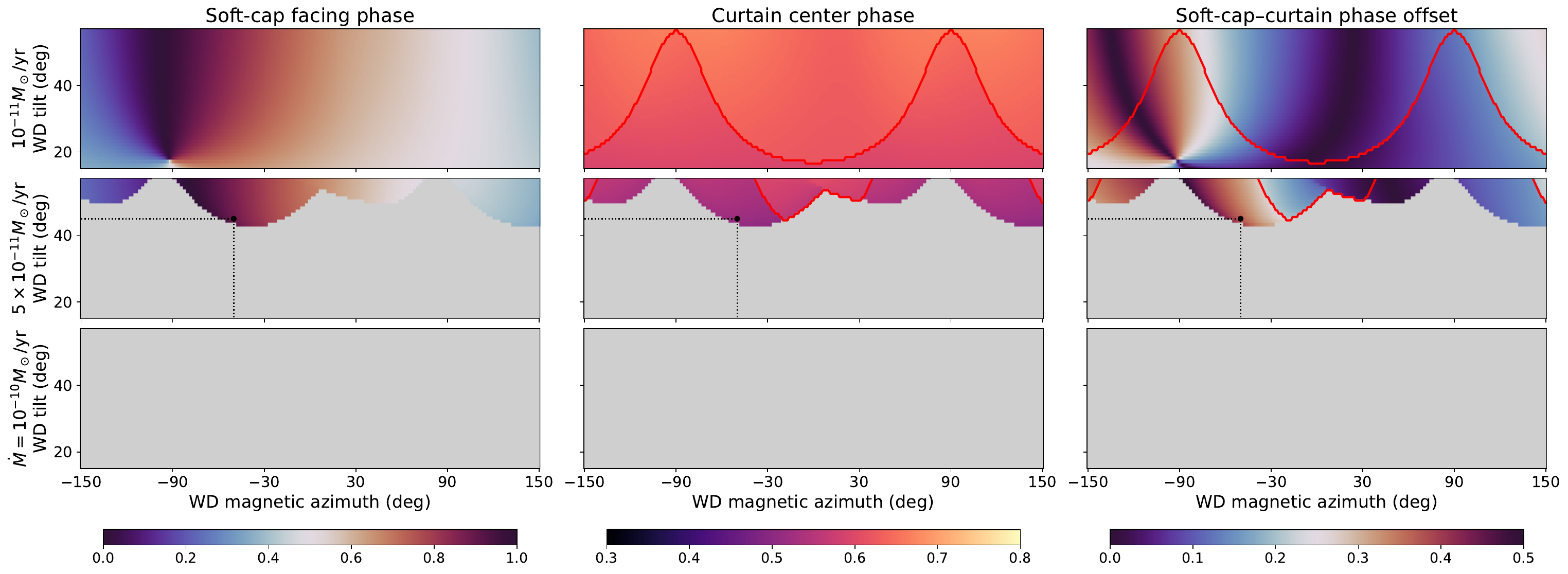}
    \caption{Dependence of the X-ray phase geometry on WD dipole orientation and mass-transfer rate. From top to bottom, the rows show $\dot M=10^{-11}$, $5\times10^{-11}$, and $10^{-10}~M_\odot\,{\rm yr^{-1}}$. The horizontal and vertical axes give WD magnetic azimuth and tilt. Columns show the soft-cap facing phase, loading-curtain centre phase, and soft-cap--curtain phase offset, respectively. Gray denotes configurations that do not satisfy the adopted full-capture condition before the first periapsis. The black point and dotted guides mark the fiducial configuration at $\dot M=5\times10^{-11}~M_\odot\,{\rm yr^{-1}}$, tilt $45^\circ$, and azimuth $-50^\circ$. The red curves in the middle and right columns separate two loading regimes: above the curves, the stream is loaded onto the magnetic field directly at L1, whereas below them, loading occurs along the stream between the WD and MD.}
    \label{fig:wd_mag_scan}
\end{figure*}

 The WD dipole orientation and mass-transfer rate determine the loading geometry. Figure~\ref{fig:wd_mag_scan} surveys WD magnetic azimuth and tilt for $\dot M=10^{-11}$, $5\times10^{-11}$, and $10^{-10}~M_\odot\,{\rm yr^{-1}}$, with the remaining parameters fixed. The left column shows the soft-cap facing phase, the middle column shows the loading-curtain centre phase, and the right column shows the phase offset between them. The black point marks the fiducial configuration with $\dot M=5\times10^{-11}~M_\odot\,{\rm yr^{-1}}$, WD tilt $45^\circ$, and azimuth $-50^\circ$.
\par We exclude the case where the accretion stream is unable to be fully captured before the first periapsis, which is shown in gray. In this case, the uncaptured part of the stream can continue around the WD and interact with the stream that has already been captured when it is approaching its second periapsis. Analyzing the secondary interaction requires hydrodynamics, which is beyond the scope of the present model. Under this constraint, the parameter space is limited. At $\dot M=5\times 10^{-11}\,M_\odot\,{\rm yr}^{-1}$, valid solutions occur mainly at relatively large WD tilts. By contrast, no solution satisfying the full-capture condition before the first periapsis is found for $\dot M=10^{-10}\,M_\odot\,{\rm yr}^{-1}$ in the adopted domain. The red curves in the middle and right columns separate two loading regimes: above the curves, the stream is loaded onto the magnetic field directly at L1, whereas below them, loading occurs along the stream between the WD and MD.

\par Within the valid region, changing the WD magnetic azimuth or tilt shifts the loading points and the connected footprint. These shifts alter the relative phases of cap projection $\mathcal V_{\rm s}$ and curtain absorption $\mathcal T_{\rm s}$ (illustrated in Sect. \ref{sec:loading_framework}). The fiducial point provides one geometry for the observed modulation, while the scan displays the variation of its underlying phase relations.

\par The fiducial parameters adopted in the main context therefore represent one viable configuration for the observed X-ray pattern, rather than a unique determination of the WD magnetic orientation. The pattern is expected to vary with azimuth and tilt because both the emitting footprint and absorbing curtain change. 

\bibliography{FRB-GW}{}
\bibliographystyle{aasjournal}

\end{document}